\documentclass[11pt,preprint]{aastex}
\usepackage{graphicx}
\usepackage{txfonts}

\newcommand{\be}{\begin{equation}}	
\newcommand{\ee}{\end{equation}}

\def\simge{\mathrel{\rlap{\raise 0.511ex
     \hbox{$>$}}{\lower 0.511ex \hbox{$\sim$}}}}
\def\simle{\mathrel{\rlap{\raise 0.511ex
      \hbox{$<$}}{\lower 0.511ex \hbox{$\sim$}}}}
\newcommand{\gae}{\lower 2pt \hbox{$\, \buildrel {\scriptstyle >}\over {\scriptstyle
\sim}\,$}}
\newcommand{\lae}{\lower 2pt \hbox{$\, \buildrel {\scriptstyle <}\over {\scriptstyle
\sim}\,$}}
\begin{document}

\title{Constraining the Symmetry Parameters of the Nuclear Interaction}
\author{James M.\ Lattimer}
\affil{Department of Physics and Astronomy, Stony Brook
University, Stony Brook, NY 11794-3800, USA}
\and
\author{Yeunhwan Lim}
%\affil{Department of Physics and Astronomy, Stony Brook
%University, Stony Brook, NY 11794-3800, USA}
\affil{Department of Physics Education, Daegu University, Gyeongsan 712-714, Korea}
\email{james.lattimer@stonybrook.edu\\ yeunhwan.lim@gmail.com}

\begin{abstract}
  One of the major uncertainties in the dense matter equation of state
  has been the nuclear symmetry energy.  The density dependence of the
  symmetry energy is important in nuclear astrophysics, as it controls
  the neutronization of matter in core-collapse supernovae, the radii
  of neutron stars and the thicknesses of their crusts, the rate of
  cooling of neutron stars, and the properties of nuclei involved in
  r-process nucleosynthesis.  We show that fits of nuclear masses to
  experimental masses, combined with other experimental information
  from neutron skins, heavy ion collisions, giant dipole resonances
  and dipole polarizabilities, lead to stringent constraints on
  parameters that describe the symmetry energy near the nuclear
  saturation density.  These constraints are remarkably consistent
  with inferences from theoretical calculations of pure neutron
  matter, and, furthermore, with astrophysical observations of neutron
  stars.  The concordance of experimental, theoretical and
  observational analyses suggests that the symmetry parameters
    $S_v$ and $L$ are in the range 29.0--32.7 MeV and 40.5--61.9 MeV,
    respectively, and that the neutron star radius, for a 1.4
    M$_\odot$ star, is in the narrow window 10.7 km $< R <$ 13.1
    km ($90\%$ confidence).  We can also set tight limits to the size
    of neutron star crusts and the fractional moment of inertia they
    contain, as well as the overall moment of inertia and quadrupole
    polarizability of 1.4 M$_\odot$ stars. Our results also have
    implications for the disk mass and ejected mass of compact mergers
    involving neutron stars. 
\end{abstract}


%\pacs{26.60.Kp, 97.60.Jd, 21.65.Cd}

\maketitle

\section{Introduction} 
The nuclear symmetry energy, and, in particular, its density
dependence, has been the focus of much recent activity due to its
importance to astrophysics~\citep{LP07}.  The expected neutrino signal
from gravitational collapse supernovae within our Galaxy is sensitive
to the symmetry energy both before core bounce~\citep{SLM94} and at
later times during the proto-neutron star stage~\citep{Roberts12}.  The
symmetry energy near $n_s$, the baryon density at saturation,
determines the radii~\citep{LP01} of neutron stars.  Neutron star radii
strongly influence the expected gravitational signals from mergers
involving neutron stars~\citep{Bauswein12,Lackey12}.  The symmetry energy also
affects the crust's thickness and thermal relaxation time, observable
in cooling and accreting~\citep{PR12} neutron stars and in giant
magnetar flares~\citep{Hurley05, TD01, SA07}. The composition of matter at
densities above $n_s$, and the existence of neutrino processes which
can rapidly cool neutron stars, depend on the symmetry
energy~\citep{LPPH91}, as do predicted properties of neutron-rich
nuclei and reaction rates involved in the astrophysical
r-process~\citep{Nikolov11}.

There have been many experimental attempts to constrain the nuclear
symmetry energy, including measurements of neutron skin thicknesses,
dipole polarizabilities, giant and pygmy dipole resonance energies,
flows in heavy-ion collisions, and isobaric analog states (see, 
for example, recent summaries by \cite{Tsang12,Newton11}).  These
constraints are weakened by varying degrees of model dependence.  Often
overlooked, however, is the important relatively model-independent
constraint arising from fitting nuclear masses, which comprise the
most abundant and accurate data available.  In combination with other
experimental data, we show that including this mass constraint leads to tight bounds which are also
compatible with astrophysical observations of neutron stars and
theoretical studies of neutron matter.

\section{The Nuclear Symmetry Energy}
The nuclear symmetry energy, $S(n)$, is the difference between the
bulk energy per baryon ($e(n,x)$) of pure neutron ($x=0$) and isospin
symmetric matter ($x=1/2$) at baryon density $n$.  We consider
baryonic matter consisting of neutrons and protons only, and $x$ is
the proton fraction.  A related quantity is the density-dependent
quadratic coefficient, $S_2(n)$, of an expansion of the bulk energy
per baryon in the neutron excess $1-2x$:
\begin{equation}
e(n,x)=e(n,1/2)+S_2(n)(1-2x)^2 + \dots
\label{symm}
\end{equation}
A common approximation is to retain only the quadratic term in
Equation \ref{symm}, even for small proton fractions $x$, so that
$S_2(n)\simeq S(n)$.  For matter in nuclei, this approximation is
valid, but care has to be taken at densities in excess of $n_s$, where
the effects of higher-order-than-quadratic terms are
amplified~\citep{Steiner06}.  Depending on the nuclear interaction,
this can be important for comparisons with neutron star matter and
with pure neutron matter.

It is customary to introduce the symmetry energy parameters
\begin{eqnarray}
S_v&=&S_2(n_s),\cr 
L&=&3n_s(dS_2/dn)_{n_s},\cr 
K_{sym}&=&9n_s^2(d^2S_2/dn^2)_{n_s},\cr
Q_{sym}&=&27n_s^3(d^3S_2/dn^3)_{n_s}.
\label{symm1}
\end{eqnarray}
Although there are limited constraints
on $K_{sym}$ from measurements of the giant isoscalar monopole
resonance~\citep{Li10}, isotopic transport ratios in medium-energy
heavy ion collisions~\citep{Chen07}, and neutron skin data from
anti-protonic atoms~\citep{Centelles09}, the extraction of the symmetry
contribution to the total incompressibility of nuclei and its
separation into bulk ($K_{sym}$) and surface contributions is still
problematic.  There are no experimental constraints on $Q_{sym}$.  
In contrast, there are several possible experimental
constraints on values of $S_v$ and $L$.

\section{Nuclear Mass Fitting} 
The most accurate and least ambiguous constraint comes from nuclear
masses.  \cite{Farine78} and \cite{Oyamatsu03} showed that fitting
Hartree-Fock and Thomas-Fermi nuclear models to mass data leads to a
linear correlation between $S_v$ and $L$, but they did not explore its
statistical significance.  \cite{Lattimer92,Lattimer96} and
\cite{D03} performed simplified statistical analyses utilizing the
liquid droplet model.  However, a more rigorous analysis was
recently undertaken by the UNEDF collaboration~\citep{Bertsch07} using
a sophisticated microscopic nuclear energy density functional
\citep{Kortelainen10}.  Before discussing their results, it is
illustrative to infer the nature of this correlation through a
two-step procedure involving the liquid drop model for nuclei and the
properties of the nuclear surface.  This illustration also
demonstrates that the correlation between symmetry parameters
determined from nuclear mass fits is not very model-dependent.

\subsection{Liquid Drop Model Illustration}
The standard liquid drop model~\citep{Myers66} holds that the
energy of a nucleus is approximately
\begin{eqnarray} E(Z,A)&=&A(-B+S_vI^2)+A^{2/3}(E_s-S_sI^2)~+\cr
&+&{3\over5}{e^2Z^2\over
r_0A^{1/3}}+E_{shell}(Z,A-Z)+E_{pairing}(A),\label{ld} 
\end{eqnarray} 
where $B\simeq16$ MeV is the bulk binding energy, per baryon, of
symmetric matter, $E_s\simeq19$ MeV is the symmetric matter
surface energy parameter.  $I=(A-2Z)/A$ is the neutron excess, and
$r_0=(4\pi n_0/3)^{-1/3}$ where $n_0$ is the nuclear saturation
density.  The last three terms in Eq. (\ref{ld}) represent the
Coulomb, shell and pairing energies, respectively.

The net symmetry energy of an isolated nucleus is then \be
E_{D,i}=I_i^2(S_vA_i-S_sA_i^{2/3}),\label{ldsym} \ee so a linear
correlation between $S_v$ and $S_s$ is expected from minimizing the
differences between model predictions and experimentally measured
symmetry energies, i.e., minimizing \be
\chi^2=\sum_i(E_{exp,i}-E_{D,i})^2/({\cal N}\sigma_D^2),\label{chi}
\ee where ${\cal N}$ is the total number of nuclei and $\sigma_D$ is a
nominal error.  A $\chi^2$ contour one unit above the minimum value
represents the $1-\sigma$ confidence interval which is an ellipse in this
linear example.

The properties of the confidence ellipse are determined by the second
derivatives of $\chi^2$ at the minimum,
\begin{eqnarray}
[\chi_{vv},~\chi_{vs},~\chi_{ss}]&=&{2\over{\cal N}\sigma_D^2}\sum_iI_i^4[A_i^2,~-A_i^{5/3},~A_i^{4/3}]\cr
&\simeq&[61.6,~-10.7,~1.87],\nonumber
\end{eqnarray}
where $\chi_{vs}=\partial^2\chi^2/\partial S_v\partial S_s$, etc.  The
specific values follow from the set of 2336 nuclei with $N$ and $Z$
greater than 40 included in the \cite{audi03} table.  The confidence
ellipse in $S_s-S_v$ space has orientation
$\alpha_{D}=(1/2)\tan^{-1}|2\chi_{vs}/(\chi_{vv}-\chi_{ss})|\simeq9.8^\circ$
with respect to the $S_s$ axis, with error widths
$\sigma_{v,D}=\sqrt{(\chi^{-1})_{vv}}\simeq2.3\sigma_D$ and
$\sigma_{s,D}=\sqrt{(\chi^{-1})_{ss}}\simeq13.2\sigma_D$ where $(\chi^{-1})$
is the matrix inverse.  The correlation coefficient is
$r_D=\chi_{vs}/\sqrt{\chi_{vv}\chi_{ss}}\simeq0.997$.  In this simple
example, the shape and orientation of the confidence interval depend
only on $A_i$ and $I_i$ and not the location of the $\chi^2$ minimum
or other drop parameters.

Although the specific values of the best-fit parameters and the exact
correlation are sensitive to the inclusion of terms representing
shell, pairing, Coulomb diffusion, exchange and Wigner energies,
deformations, and the neutron skin, among others, their effects are
relatively minor.  For example,
\cite{van06} showed that the Wigner term is important in identifying
the best-fit values of the liquid drop parameters, but does not
significantly affect the deduced correlation between $S_v$ and $S_s$.

\subsection{Liquid Droplet Model}
In practice, the liquid droplet model~\citep{Myers69}, which differs
from the liquid drop model by accounting for varying neutron/proton
ratios within the nucleus that produce neutron skins in neutron-rich
nuclei, is an improved treatment.  Its symmetry energy is
\be
E_{LD,i}=S_vI_i^2A_i(1+S_sA_i^{-1/3}/S_v)^{-1},
\label{ld1}
\ee 
and therefore predicts a
linear correlation between $S_s/S_v$ and $S_v$ rather than between
$S_s$ and $S_v$ as in the drop model. 
 The same methodology as for the liquid
drop model can be used to determine the confidence interval in
$S_s/S_v - S_v$ space.  In addition to $A_i$ and $I_i$, the properties
of the confidence interval now depend, to a degree, on the measured
masses, $E_{exp,i}$, as well as the parameters of the droplet model.

The correlation between $S_s/S_v$ and $S_v$ emerging from nuclear mass
fitting is roughly equivalent to a correlation between $L$ and $S_v$
because the surface symmetry parameter, $S_s$, depends on the behavior
of $S(n)$ within the nuclear interior.  The general features can be
appreciated using the plane-parallel approximation for the nuclear
surface.  At zero temperature, for densities less than the nuclear
saturation density $n_s$, bulk matter separates into two phases of
differing densities and proton fractions (a dense phase and a dilute
phase) having equal pressures and neutron and proton chemical
potentials.  The surface thermodynamic potential per unit area,
$\sigma$, also known as the surface tension, is the difference between
the total thermodynamic potential and that of two uniform fluids
each having the properties of matter far from the interface on both
sides.  It is straightforward to show that in the case appropriate for
finite nuclei, in which the density in the dilute phase ($z=+\infty$)
vanishes, \be \sigma=2\int_{-\infty}^{+\infty}n[e(n,x)-(1-x)\mu_n
-x\mu_p]dz,
\label{sigma} 
\ee 
where $\mu_n$ and $\mu_p$ are the neutron and proton chemical
potentials, respectively.  To lowest order in the dense phase neutron
excess $I=1-2x_{-\infty}$, the surface tension can be expanded: 
\be
\sigma\simeq\sigma_0-(1-2x_{-\infty})^2\sigma_x,
\label{sigma1}
\ee where $\sigma_0$ is the symmetric matter surface tension.  In this
notation, the surface energy parameters of the liquid drop and droplet
models are $E_s=4\pi r_0^2\sigma_0$ and $S_s=4\pi r_0^2\sigma_x$.  

One can show that to leading order~\citep{LS89,D03,SPLE05} 
\be
{S_s\over S_v}={E_s\over2}{\int_0^{1}u^{1/2}[S_v/S_2(u)-1]f(u)^{-1/2}
du\over\int_0^1u^{1/2}f(u)^{1/2}du},\label{sssv}
\ee 
where $u=n/n_s$ and $f(u)=e(u,1/2)+B$.  A simple analytical
relation is found if we approximate $S_2(u)\simeq S_v+L(u-1)/3$ and
$f(u)\simeq K_0(u-1)^2/18$, where $K_0\simeq240$ MeV is the
incompressibility parameter:
\begin{equation}
{S_s\over S_v}\simeq{135E_s\over2K_0}\left[1-\eta\tan^{-1}\eta^{-1}\right],\label{sssv1}
\end{equation}
where $\eta=\sqrt{3S_v/L-1}$.  Assuming $L/S_v=1,3/2$ or $2$, one
finds, respectively, $S_s/S_v\simeq0.7,1.2$ or $1.8$, which shows the
basic trend although the values are somewhat less than given by realistic
calculations.  

This illustration is model-dependent, partly because it is based on
the liquid droplet model and the semi-infinite surface approximation,
but mostly because of the highly simplified assumed dependence of
$S(n)$ on $L$ and $S_v$.  In actuality, $S(n)$ depends on higher order
terms such as $K_{sym}$ and $Q_{sym}$.  We have found, applying the
finite-range nuclear interaction~\citep{MS90} to Thomas-Fermi
semi-infinite surface calculations, that $S_s/S_v$ can be approximated
by \be\label{sssv2} {S_s\over
  S_v}\simeq0.6461+{S_v\over97.85{\rm~MeV}}+ 0.4364{L\over
  S_v}+0.0873\left({L\over S_v}\right)^2 \ee with a mean error of
0.6\%.  We found that a very similar relation emerges from
Thomas-Fermi calculations using generalized Skyrme functionals.  Very
approximately, $S_s/S_v$ increases linearly with $L$, so that one
expects the confidence interval in $L-S_v$ space to have a highly
elongated elliptical shape with a small angle with respect to the $L$
axis.

\begin{figure}[h]
\vspace*{-.8cm}
\hspace*{-1.3cm}
\includegraphics[angle=90,width=14cm,angle=90]{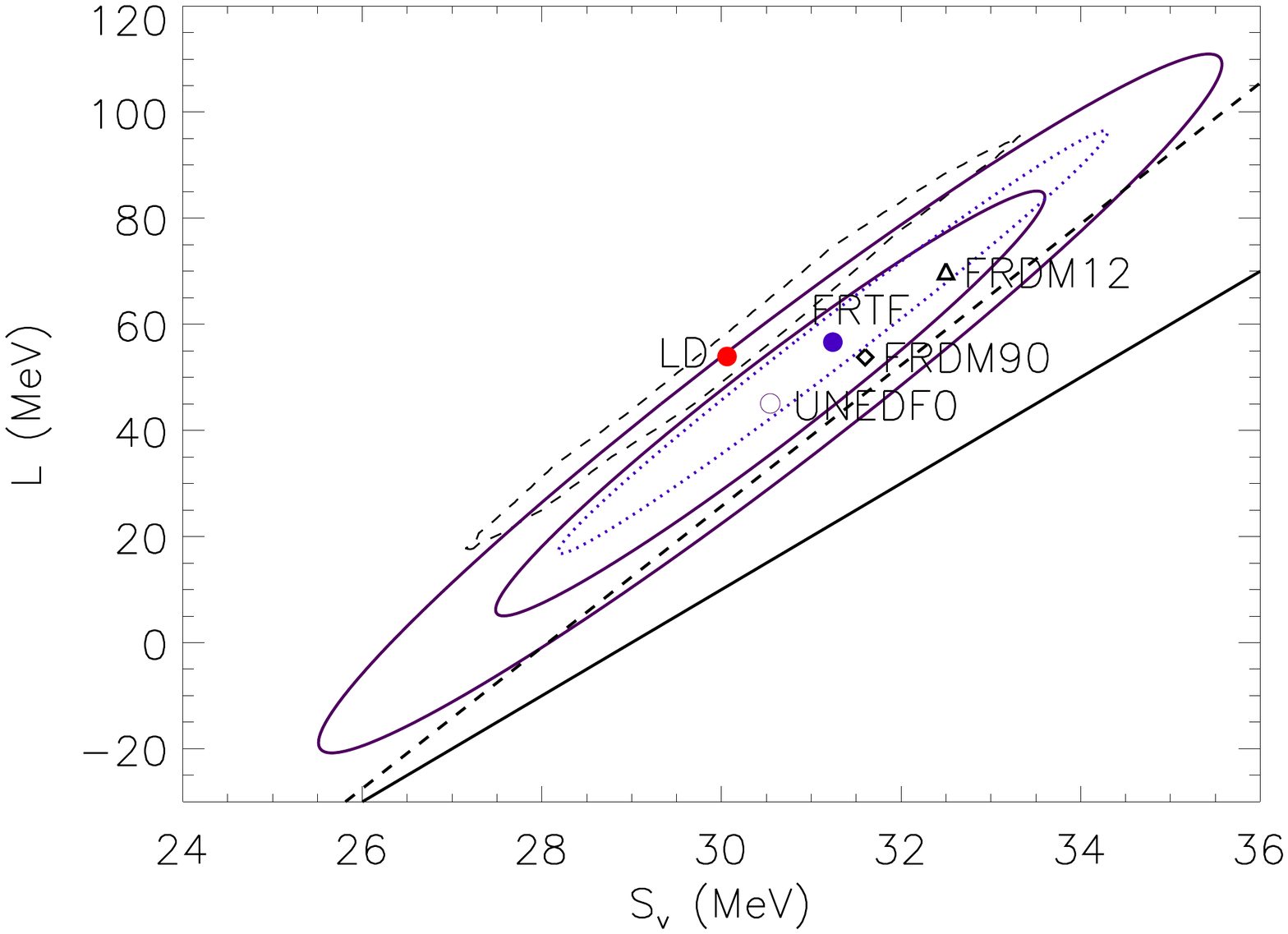}
%\vspace*{-1.5cm}
\caption{Comparison of confidence intervals for nuclear mass fitting.
  Solid figures are the UNEDF0 68\% and 95\% confidence intervals of
  \cite{Kortelainen10} assuming $\sigma_{UNEDF0}=2$ MeV.  The dashed
  and dotted figures denote the 68\% confidence intervals for a
  liquid droplet fit assuming $\sigma_{LD}=0.5$ MeV and a Thomas-Fermi
  finite-range fit assuming $\sigma_{FRTF}=1.6$ MeV, respectively.
  Circles mark values of $S_{v}$ and $L$ at the respective $\chi^2$ minima.
  The solid line is the correlation of
  \cite{Farine78} and the dashed line is the correlation of
  \cite{Oyamatsu03}.  The diamond~\citep{MS90} and
  triangle~\citep{Moller12} show finite-range liquid droplet mass
  fits.}
\label{massfit}
\end{figure}

Least-squares fitting of a liquid droplet model, together with the
Thomas-Fermi result in Eq.~\ref{sssv2}, leads to best-fit values of
the symmetry parameters $S_{v0,LD}=29.1$ MeV and $L_{0,LD}=53.9$ MeV.
Shell effects were modeled with linear and quadratic terms in the
neutron and proton valence numbers (the smallest distances from their
respective magic numbers) as in \cite{van06}.  Pairing effects were
modeled with a term $a|A|^{1/2}$ for odd-odd nuclei, $-a|A|^{1/2}$ for
even-even nuclei, and zero for odd-even nuclei, where $a\sim12$ MeV
was found from the fitting procedure.  Conventional Wigner and Coulomb
exchange energy terms were also employed, as was a further term
representing the Coulomb diffuseness energy. The inclusion of these auxiliary terms, while
improving the overall fit, which was $\chi_{min}\sigma_{LD}\simeq1.26$
MeV, did not significantly alter the inferred correlation between
$S_v$ and $L$, although the location of the $\chi^2$ minimum does
depend on them.  The results scale
  with the assumed value of the nominal error, in this case
  $\sigma_{LD}$.  We also found that the properties of the confidence
interval were$\sigma_{v,LD}=6.2\sigma_{LD}$,
$\sigma_{L,LD}=80.0\sigma_{LD}$ and $\alpha_{LD}=15.3^\circ$
(Figure~\ref{massfit}).  In spite of the simplified nature of this
model, its predictions turn out to be robust when compared to those of
more sophisticated microscopic models.

\subsection{Mass Fit Correlations from Microscopic Nuclear Models}

The fitting of nuclear interaction parameters to binding
energies using microscopic nuclear models was undertaken by
\cite{Farine78}.  Using a number of different non-relativistic
effective interactions, they performed a series of Hartree-Fock
calculations and found a nearly linear correlation between $L$ and
$S_v$: $L\simeq10S_v-290$ MeV (Figure~\ref{massfit}).  Later,
\cite{Oyamatsu03} showed that fitting non-relativistic and
relativistic Hartree-Fock, as well as constrained variational
Thomas-Fermi, nuclear models to mass data lead to the linear
correlation $L\simeq13.3S_v-373.3$ MeV (Figure~\ref{massfit}).  Neither
publication explored the statistical significance of their results.
The slope of the \cite{Oyamatsu03} correlation matches that of the 
liquid droplet model, but that of the \cite{Farine78} correlation
is discrepant.

\cite{Kortelainen10}, simultaneously fitting nuclear masses and charge
radii with the UNEDF0 potential and the Hartree-Fock methodology,
explored the significance of the symmetry energy correlation.  They
obtained $S_{v0,UNEDF0}\simeq30.5$ MeV, $L_{0,UNEDF0}\simeq45.1$ MeV,
$\sigma_{v,UNEDF0}\simeq1.58\sigma_{UNEDF0}$ MeV,
$\sigma_{L,UNEDF0}\simeq20.0\sigma_{UNEDF0}$ MeV and
$\alpha\simeq15.2^\circ$.  \cite{Kortelainen10} adopted the somewhat
arbitrary value of $\sigma_{UNEDF0}=2$ MeV, resulting in the
68\% and 95\% confidence intervals shown in Figure \ref{massfit}.
The RMS error of the UNEDF0 fit was $\chi\sigma_{UNEDF0}=1.45$ MeV.

This result closely matches fits obtained with both liquid droplet and
Thomas-Fermi nuclear models (Figure~\ref{massfit}).  The length of the
confidence interval for the liquid droplet model has the same size if
we choose $\sigma_{LD}=0.50$ MeV.  For the finite-range Thomas-Fermi
(FRTF) fits, the parameters for the shell, pairing, Wigner and Coulomb
exchange energies were taken from the liquid droplet best-fit model
rather than including them in the fitting procedure.  We obtained
$\chi\sigma_{FRTF}\simeq2.75$ MeV, $S_{v0,FRTF}\simeq31.2$ MeV and
$L_{0,FRTF}\simeq56.6$ MeV.  The length of the Thomas-Fermi
finite-range confidence interval equals that of ~\cite{Kortelainen10}
if $\sigma_{FRTF}=3.1$ MeV.

The correlation coefficient between $S_v$ and $L$ is approximately
0.97 for UNEDF0, less than the nearly unity value estimated from the
liquid droplet or Thomas-Fermi models since the UNEDF0 model was also
constrained by nuclear radii.  It should be noted that the 95\%
confidence interval contains symmetry parameter values that may be
somewhat unphysical from the point of view of the properties of
neutron-rich matter near and above the saturation density.  For
example, negative values of $L$ imply that pure neutron matter has
negative pressures near $n_s$, implying stability for neutron
clusters.  The size of the confidence ellipse scales with the assumed value of
$\sigma_{UNEDF0}$, so had a value $\sigma_{UNEDF0}=1$ MeV been chosen
instead, the 95\% confidence interval would have extended no lower than
$L\simeq20$ MeV.  

Another measure that can be applied to estimate
the sizes of confidence intervals is the RMS error of the fit.  
For example, the liquid droplet 68\% confidence interval displayed
in Figure~\ref{massfit} corresponds to an RMS error of approximately
0.013 MeV/baryon.  In other words, applied to liquid drop dense matter equations
of state, such as that of \cite{LS91}, symmetry parameters chosen from within
this confidence region would yield errors in neutron or proton chemical
potentials of less than about 0.013 MeV (aside from those arising from neglected shell and pairing effects).

The UNEDF collaboration has published a second parameter set, UNEDF1
\citep{Kortelainen12}, for a modified energy density functional
suitable for studies of strongly elongated nuclei.  In this study,
binding energies and charge radii, as well as excitation energies, of spherical nuclei, deformed nuclei
and fission isomers, were fit.  The best-fit values for the symmetry parameters of
UNEDF1 lie near the $1-\sigma$ confidence ellipse of UNEDF0, as shown
in Figure 1.  However, this study did not discuss a correlation
between $S_v$ and $L$.  Because this fit is heavily weighted by deformed nuclei, and
because the fit includes excitation energies, it will not show the same trends as the UNEDF0 fit.

As a final demonstration that this correlation is robust, the
symmetry parameters determined from nuclear mass-fitting
with the finite-range droplet model (FRDM)~\citep{MS90,Moller12} has
consistently produced values of $S_v$ and $L$ along the UNEDF0
correlation line (Figure~\ref{massfit}).  This occurs despite
differences among nuclear models, including the algebraic forms of the
nuclear interaction and treatments of shell, pairing, deformation and
Wigner contributions.  Unfortunately, confidence intervals for the FRDM
fits were not published.

The correlation between symmetry parameters established from nuclear
mass fitting has previously been used to constrain symmetry energy
variations in supernova simulations~\citep{SLM94}, but,
surprisingly, has rarely been combined with results derived from
other experimental data.  We show that doing so considerably enhances
the constraints established from other experiments.

\section{Other Experimental Constraints} The liquid droplet model
implies that other nuclear observables will be related to the surface
and volume symmetry coefficients and should provide additional
constraints on them.  Some of these constraints were summarized by \cite{Newton11,Tsang12}.

\subsection{Neutron Skin Thickness}
An example is the neutron skin thickness of neutron-rich nuclei.
Neglecting Coulomb effects, the difference between the mean neutron
and proton surfaces in the liquid droplet model~\citep{Myers69} is
\be\label{skin} t_{np}={2r_o\over3}{S_sI\over S_v+S_sA^{-1/3}}, \ee
which is, to lowest order, a function of $S_s/S_v$.  Given that
$S_s/S_v$ is largely a function of $L$, it should thus form a nearly
orthogonal constraint to that from nuclear masses.  Values of $t_{np}$
have been measured, typically with 30-50\% errors, for a few dozen
nuclei.  More frequently used is the neutron skin thickness $r_{np}$, 
the difference between the mean square neutron and proton radii, which is
approximately equal to $\sqrt{3/5}t_{np}$.

\begin{figure}[h]
%\vspace*{-1.cm}
%\hspace*{-.6cm}
\includegraphics[angle=270,width=18cm,angle=90]{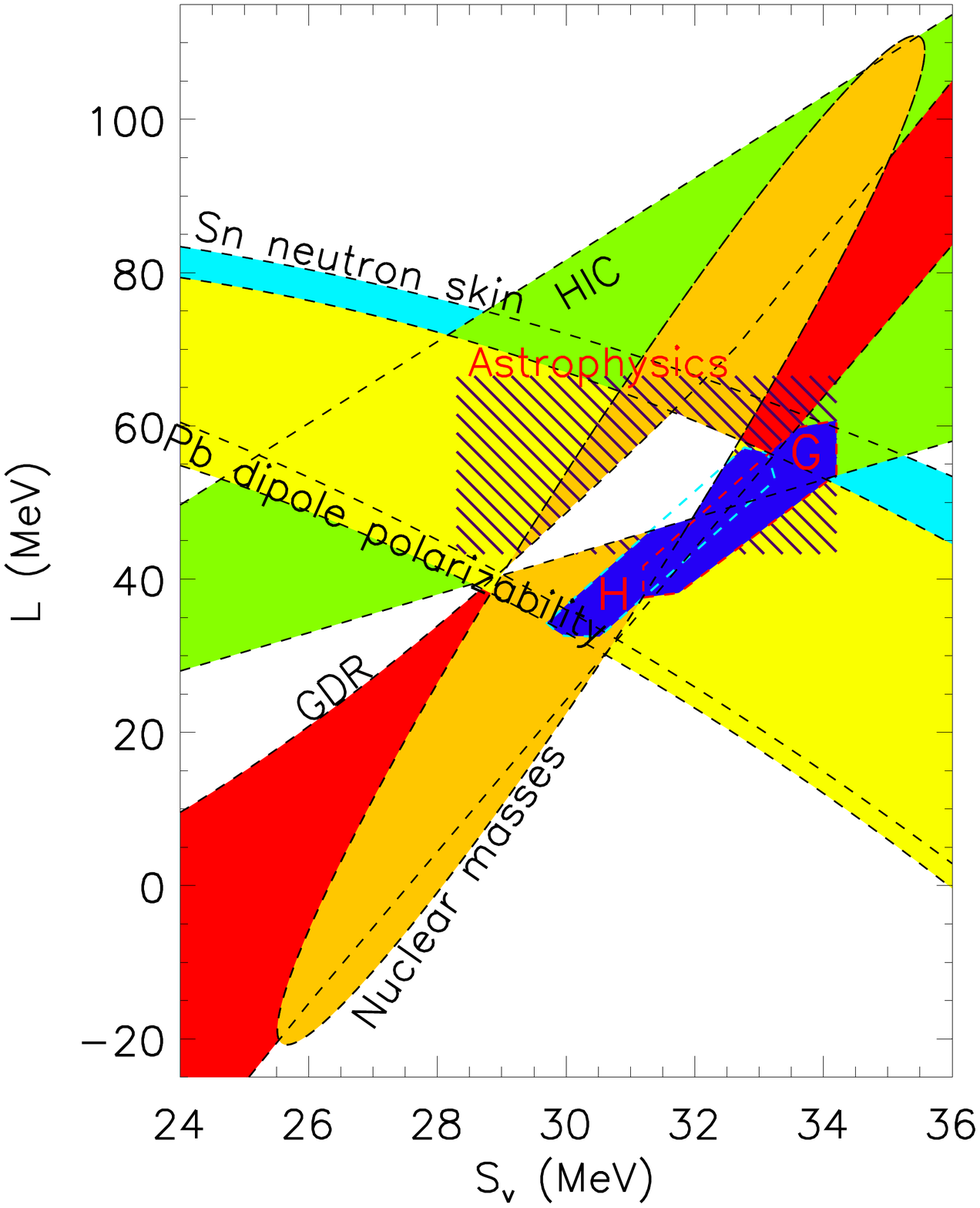}
%\vspace*{-1cm}
\caption{Summary of constraints on symmetry energy parameters.  The
  filled ellipsoid indicate joint $S_v-L$ constraints from nuclear
  masses~\citep{Kortelainen10}.  Filled bands show constraints from
  neutron skin thicknesses of Sn isotopes~\citep{Chen10}, the dipole
  polarizability of $^{208}$Pb~\citep{Piekarewicz12}, giant dipole
  resonances (GDR)~\citep{Trippa08}, and isotope diffusion in heavy ion
  collisions (HIC)~\citep{Tsang09}.  The hatched rectangle shows constraints
  from fitting astrophysical $M-R$ observations~\citep{SLB10,SLB13}.  The two
  closed regions show neutron matter constraints (H is
  \cite{HLPS10} and G is \cite{GCR12}).  The enclosed  white area is
  the experimentally-allowed overlap region.}
\label{correl}
\end{figure}

A recent study of Sn isotopes~\citep{Chen10}, where differential
isotopic measurements with fixed $Z$ potentially reduce errors,
produced a correlation band in the $S_v-L$ plane which is nearly
orthogonal to mass-fit correlations (Figure~\ref{correl}).  The
results from the Sn isotopes can, approximately, be expressed as a range of
values of the neutron skin thickness of $^{208}$Pb, $r_{np}^{208}$, since
numerous theoretical studies have shown them to be highly correlated
(e.g., \cite{RN10}).

A number of authors have demonstrated from Hartree-Fock and
Thomas-Fermi studies that $r_{np}^{208}$ varies linearly with $L$ for
a given value of $S_v$ (see, for example, \cite{furnstahl02}).
However, $r_{np}^{208}$ can also be shown to vary linearly with $S_v$
if $L$ is fixed \citep{furnstahl02,RN10,Chen10}.  In general,
$r_{np}^{208}$ is a function of both $S_v$ and $L$, and the high
degree of correlation between $S_v$ and $L$ is responsible for these
separate, nearly linear, correlations.

\cite{Chen10} established a two-dimensional relation,
$r_{np}^{208}(S_v,L)$, using Skyrme Hartree-Fock calculations of
$^{208}$Pb in which most Skyrme parameters were determined from
nuclear matter properties, and $S_v$ and $L$ were then systematically
varied.  We can approximate their results with \be\label{rnschen}
r_{np}^{208}(S_v,L)\simeq a_0+a_1S_v+a_2L+a_3S_v^2+a_4S_vL+a_5L^2, \ee
where the coefficients are given in Table~\ref{tableskin}.  The mean
error of this fit is about 0.7\% over the ranges of interest.  We
attempted to reproduce the results for $r_{np}^{208}(S_v,L)$ with a
Skyrme Hartree-Fock analysis, and found similar results to within
about 0.01 fm over the relevant ranges of $S_v$ and $L$.  The
coefficients of our fit are also given in Table~\ref{tableskin}.  
  The fits are not identical due to differences in the assumed energy
  density functionals as well as errors in extracting values from
  Figure 7 of \cite{Chen10}. By combining the linear
correlation between $L$ and $S_v$ from \cite{Kortelainen10}, namely
$L\simeq-348.45{\rm~MeV}+12.903S_v$, with Eq.~(\ref{rnschen}), we are
able to reproduce the approximate relations between $r_{np}^{208}$ and
$L$ found by \cite{furnstahl02,chen05}; and \cite{Centelles09}.

\cite{B00} and \cite{TB01} found an approximately linear correlation between $r_{np}$ and the
pressure ($p_N(0.1)$) of pure neutron matter at the density $0.1$ fm$^{-3}$, 
\begin{equation}\label{tb}
r_{np}=(0.060\pm0.015)+0.12p_N(0,1).
\end{equation}  
This correlation is phenomenological because matter in the neutron-rich nuclear surface is very different from pure neutron matter.
$p_N(0.1)$ cannot be completely specified by $S_v$ and $L$, partly because of variations in the symmetric matter pressure and higher order terms in the expansion of $S_2$ with respect to density (see \S~\ref{sec:crust}).  Nevertheless, we have verified that Eq.~(\ref{rnschen}), with the parameters in Table~\ref{tableskin}, reproduces this correlation.
 
\begin{deluxetable}{lcc}
\tablecolumns{3}
\tablewidth{0pc}
\tablecaption{Fitting coefficients for $r_{np}^{208}$ in Eq.~(\ref{rnschen}).\label{tableskin}}
\tablehead{
        \colhead{Coefficient} & \colhead{\cite{Chen10}} & \colhead{This paper}
}
\startdata
$a_0$ & -0.094669 & -0.11767\\
$a_1$ & 0.0072028 & 0.0070281\\
$a_2$ & 0.0023107 & 0.0027521\\
$a_3$ & $-8.8453\cdot10^{-6}$ & $4.7517\cdot10^{-7}$\\
$a_4$ & $-4.7837\cdot10^{-5}$ & $-6.0221\cdot10^{-5}$\\
$a_5$ & $4.0030\cdot10^{-6}$ & $4.9664\cdot10^{-6}$
\enddata
\end{deluxetable}

The limits on $r_{np}^{208}$ deduced from the neutron skin thickness
of Sn isotopes by \cite{Chen10}, $0.155{\rm~fm}\simle
r_{np}^{208}\simle0.195{\rm~fm}$, yield the band shown in
Figure~\ref{correl}.  The above results indicate that this band is not
 very model-dependent.

A recent study~\citep{Krasznahorkay13} of neutron skin thicknesses in
Sn and Pb nuclei, deduced from measurements of antianalog giant dipole
resonances, yielded the constraint $0.16{\rm~fm}\simle
r_{np}^{208}\simle0.21{\rm~fm}$. \cite{Zenhiro10},
  utilizing proton elastic scattering from $^{208}$Pb, calculated
  $0.148{\rm fm}\simle r_{np}^{208}\simle0.265{\rm~fm}$.  Both agree with
the analysis of \cite{Chen10}, but neither further constrains the
symmetry parameters.  \cite{Tsang12} averaged results from various
  studies of photon emission during the decays of antiprotonic states
  of neutron-rich nuclei to deduce estimates for neutron skin
  thicknesses.  They inferred a central value $r_{np}^{208}=0.18$ fm,
  but with even larger uncertainties.  First
results~\citep{Abrahamyan12} from the PREX experiment to directly
measure the neutron skin of $^{208}$Pb were too imprecise to be
definitive, yielding $0.15{\rm~fm}\simle r_{np}^{208}\simle0.49{\rm~fm}$.  The
central value is discrepant with other results for neutron skins, but
the uncertainties are still very large.

\subsection{Giant Dipole Resonance}
The hydrodynamical droplet model of \cite{LS89}
showed that the giant dipole
resonance (GDR) centroid energy is closely connected to liquid droplet
parameters.  In this model, the centroid energy is
\begin{equation}\nonumber\label{eqgdr}
E_{-1}\simeq\sqrt{{3\hbar^2(1+\kappa)\over m<r^2>}S_v\left(1+{5\over3}{S_s\over S_v}A^{-1/3}\right)^{-1}},
\end{equation}
where $\kappa$ is an enhancement factor arising from the velocity
dependence of the interaction and $<r^2>$ is the mean-square charge
radius.  The factor $m<r^2>/(1+\kappa)$ does not significantly vary
among interactions for a given isotopic chain, and its value therefore does not
  play a role in estimating a correlation between the symmetry
  parameters.  A series of measurements of the giant
dipole resonance should result in a correlation
between $S_s/S_v$ and $S_v$ that is similar to but slightly less steep
(i.e., a greater value of $\alpha$) than that established from nuclear
masses.   Equations (\ref{ld1}) and (\ref{eqgdr})
predict that the orientation angles ($\alpha$)
of the GDR and binding energy correlations in the $S_s/S_v-S_v$ plane will have the
ratio
\begin{equation}\nonumber
{\alpha_{GDR}\over\alpha_{binding}}\approx{5\over3}\left(1+2{S_{s0}\over S_{v0}}<A^{-1/3}>\right)\left(1+{8\over3}{S_{s0}\over S_{v0}}<A^{-1/3}>\right)^{-1}\simeq1.45,
\end{equation}
where $<>$ refers to an average over the measured nuclei.  A similar ratio will then apply to the relative orientation angles in the $L-S_v$ plane.

\begin{figure}[h]
%\vspace*{-1.1cm}
%\hspace*{-.5cm}
\includegraphics[angle=270,width=18cm,angle=90]{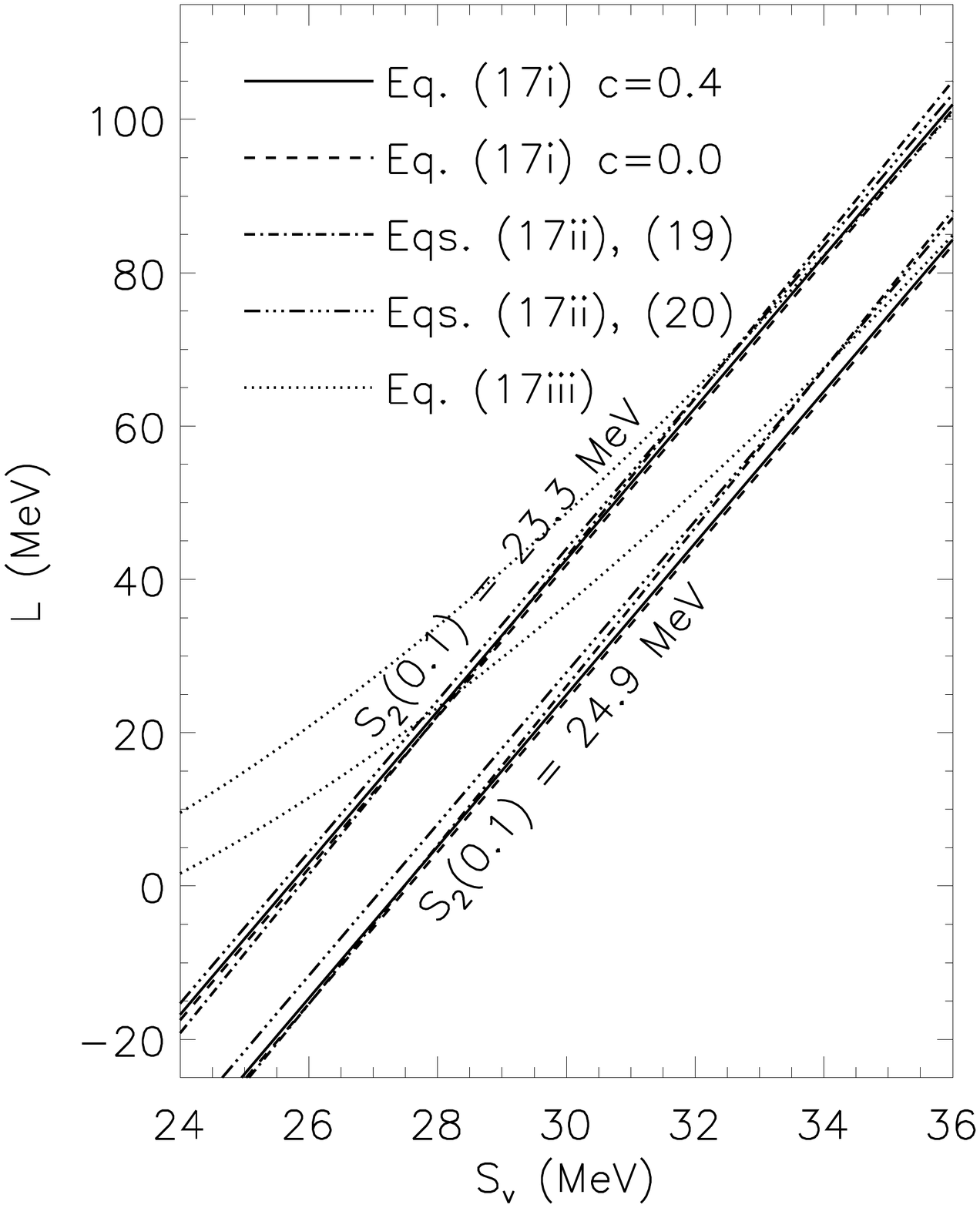}
%\vspace*{-1cm}
\caption{Bounds for symmetry energy parameters implied by the 
  constraint 23.3 MeV $<S_{2}(0.1)<$ 24.9 MeV, derived from giant
  dipole resonances~\citep{Trippa08},  combined with the various forms of
    Eq. (\ref{S20}) described in the text.}
\label{gdr}
\end{figure}

Using a variety of Skyrme forces to model the giant dipole resonance
in $^{208}$Pb, \cite{Trippa08} concluded the measured centroid energy
was best fit by those forces having a bulk symmetry energy $S_2(0.1)$,
evaluated at the density $n=0.1{\rm~fm}^{-3}$, in the
range 23.3 MeV $<S_{2}(0.1)<$ 24.9 MeV.
The relation between $S_{2}(0.1)$ and $S_v$ and $L$ is obviously model
dependent.   

To explore this model dependence, we selected a number
of functional forms for $S_2$:
\begin{equation}\label{S20}
\begin{array}{rll}
  S_2(u)&=(T_0/3)u^{2/3}(1+cu)+a_1u+b_1u^{4/3}, \qquad &(i)\cr
  S_2(u)&=S_v+(L/3)(u-1)+(K_{sym}/18)(u-1)^2+(Q_{sym}/162)(u-1)^3+d_1(u-1)^4, \qquad &(ii)\cr
  S_2(u)&=(T_0/3)u^{2/3}+[S_v-(T_0/3)]u^\gamma, \qquad &(iii)
\end{array}
\end{equation}
where $T_0=(\hbar^2/2m_b)(3\pi^2n_s/2)^{2/3}$ and
\begin{eqnarray}\label{coef}
a_1&=&4S_v-{2T_0\over3}\left(1-{c\over2}\right)-L,\qquad
b_1={T_0\over3}(1-2c)+L-3S_v,\cr
d_1&=&{L\over3}-S_v-{K_{sym}\over18}+{Q_{sym}\over162},\qquad
\gamma={3L-2T_0\over3(3S_v-T_0)}.
\end{eqnarray}
Form (i)
of Eq. (\ref{S20}) originates from Skyrme interactions.  
Here, $c$ accounts for the nucleon effective mass in the kinetic energy contribution.  We considered cases in which $c=0$ or $c=0.4$.
 Form (ii) is the
polynomial density expansion with coefficients of Eq. (\ref{symm1}).
The parameter $d_1$ ensures that $S_2(0)=0$.  To evaluate
 $K_{sym}$ and $Q_{sym}$, which are not directly
constrained by experiment, we considered two cases:  \cite{Ducoin11} argued them to be linearly
correlated with $L$ for Skyrme functionals fit to nuclear properties:
\begin{eqnarray}\label{lfitd}
K_{sym}&\simeq&3.33L-281{\rm~MeV},\cr
Q_{sym}&\simeq&-6.63L+765{\rm~MeV},
\end{eqnarray}
 while \cite{Vidana09}, using $Q_{sym}=0$, found \be\label{lfitv}
K_{sym}\simeq2.867L-260{\rm~MeV}.  \ee 
  This ensemble of four
      examples adequately describes the symmetry behavior of both
      Skyrme and relativistic field nuclear interactions.
Finally, form (iii) has often been used in fitting results from heavy-ion collisions~\citep{Zhang08}.

    The experimental constraints on $S_2(0.1)$ using these 
      variations of Eq. (\ref{S20}) are displayed in Figure
    \ref{gdr}, from which we infer that the effective model
    dependence is moderate.  For example, the terms involving
    $Q_{sym}$  and $d_1$ have little influence on these results.
    The form Eq. (\ref{S20} iii) deviates at small values of
      $S_v$.  We show in \S \ref{sec:crust} that this parameterization
      is a poor representation of neutron matter at subnuclear
      densities.  To be conservative, however, we have chosen to adopt the 
      extreme ranges to represent the experimental giant dipole
    resonance limits on $S_v$ and $L$ (Figure~\ref{correl}).
    The resulting band has a shallower slope, compared to that
    inferred from nuclear masses, as predicted.
 
\subsection{Dipole Polarizabilities}

The electric dipole polarizability, $\alpha_D$, is related to the
response of a nucleus to an externally applied electric field, and it
is largely concentrated in the giant dipole resonance.  The
approximate proportionality of $\alpha_D$ and $r_n^{208}$ was
demonstrated macroscopically by \cite{LS89} and confirmed by
microscopic calculations using energy-density functionals by
\cite{RN10}.  \cite{Tamii11} measured the dipole polarizability of
$^{208}$Pb to be $\alpha_D=(20.1\pm0.6)$ fm$^3$. Using the very high
degree of correlation between $\alpha_D$ and $r_{np}^{208}$
established by \cite{RN10}, this measurement of $\alpha_D$ is
equivalent to the range $0.131{\rm~fm}\simle
r_{np}^{208}\simle0.177{\rm~fm}$.  The analysis by \cite{RN10} was based
upon a non-relativistic density functional analysis.  A systematic
reanalysis~\citep{Piekarewicz12}, incorporating both relativistic and
non-relativistic models calibrated to nuclear properties in an
attempt to minimize model dependencies, revised this permitted range to
$0.146{\rm~fm}\simle r_{np}^{208}\simle0.190{\rm~fm}$.  This result is
nearly the same as that established from neutron skin thicknesses
of Sn isotopes (Figure~\ref{correl}).

\subsection{Heavy Ion Collisions and Further Constraints}

Constraints from heavy ion collisions include those from isospin
diffusion~\citep{Tsang09}, shown in Figure~\ref{correl}, and from
multifragmentation in intermediate-energy heavy ion
collisions~\citep{Shetty07}.  The latter implied 40 MeV $<L<$
125 MeV, which is consistent with our other constraints, but does not
further limit parameter space and is therefore not displayed in
Figure~\ref{correl} for clarity.  Both limits have been established from
specific types of nuclear interaction models, and may be subject to
considerable model dependence.

Also not included in Figure~\ref{correl} are results of a study of pygmy
dipole resonances~\citep{Carbone10} in $^{68}$Ni and $^{132}$Sn, which
implied 31 $<S_v/{\rm MeV}<$ 33.6 and 49.1 $<L/{\rm MeV}<$ 80.
\cite{DG11} argued that theoretical and experimental uncertainties
prevent them from being effective constraints.  The results of
\cite{RN10} confirmed that pygmy resonance energies lack a significant
correlation with $L$.

\cite{DL09} proposed isobaric analog states as constraints on symmetry
parameters and concluded that $S_v\simeq33$ MeV and $S_s\simeq96$ MeV,
which is equivalent to $L\approx110$ MeV.  These values were derived
from a fit including bulk and surface contributions to data for nuclei
over a broad range of masses.  However, we reanalyzed the fit
including a curvature term (proportional to $A^{-2/3}$) and determined
that $S_v\sim30$ MeV and $S_s\sim45$ MeV, which would be equivalent to
$L\approx50$ MeV.  Neither fit is statistically preferred. A more
detailed analysis will be necessary to produce better constraints.

In summary, the various experimental results jointly confine the
allowable ranges of symmetry parameters to a small region
(Figure~\ref{correl}) bounded by 29.0 MeV $<S_v<$ 32.7 MeV and 40.5
MeV $<L<$ 61.9 MeV.  The lower limit to $L$ is set by the heavy-ion
collision analysis, while the upper limit to $L$ is determined by the
dipole polarizability analysis.

\section{Neutron Matter Theory}
\label{Sec:nm}
Two recent studies of pure neutron matter have employed chiral
Lagrangian~\citep{HLPS10} and quantum Monte Carlo~\citep{GCR12}
techniques using realistic two- and three-nucleon interactions.  In
each case, errors were estimated by considering physically-motivated
ranges for interaction strengths.  Symmetry parameters can be
estimated from these calculations because, if higher-than-quadratic
contributions in Eq.\ref{symm} can be neglected, the energy and
pressure of pure neutron matter at the saturation density are
$e(n_s,0)=S_v-B$ and $p(n_s,0)=Ln_s/3$, respectively.  The horizontal
width of the permitted quantum Monte Carlo region is underestimated
because, in this calculation, $S_v$ was used as an input and was
restricted to the indicated range (31 MeV $<S_v<$ 34 MeV) based on
nuclear mass fits.  Despite their different theoretical approaches,
these independent predictions for the symmetry parameters are
remarkably consistent with each other and, furthermore, lie close to
the experimentally preferred values (Figure~\ref{correl}).

Care should be taken with the neutron matter regions plotted in
Figure~\ref{correl} because they can be translated in both $S_v$ and
$L$ if higher-than-quadratic terms exist in Eq. \ref{symm}.  In this
case the quantity \be\label{qn} Q(n)=e(n,0)-e(n,1/2)-S_2(n) \ee is
non-zero and the value inferred for $S_v$ from the neutron matter
energy at the saturation density, $e(n_s,0)$, would be smaller by
$Q(n_s)$ compared to the case in which higher-than-quadratic terms are
ignored.  The kinetic energy contribution to $Q$ for a
non-interacting, non-relativistic, degenerate Fermi gas of nucleons,
for example, is \be\label{qk}
Q_{k}(n)={3\hbar^2\over10m_b}\left(2^{2/3}-{14\over9}\right)\left({3\pi^2n\over2}\right)^{2/3}\simeq0.704\left({n\over
    n_s}\right)^{2/3}{\rm~MeV}.  \ee The shift to lower inferred
values for $L$ for this kinetic contribution would be $2Q_k(n_s)$.
Potential energy (and effective mass corrections) also contribute to
$Q$.  For many Skyrme and relativistic mean field interactions these
are negligible by construction, but the MDI (momentum-dependent
interaction) force~\citep{Das03}, often used in studies of heavy-ion
collisions, has $Q(n_s)\simeq 0.27S_2(n_s)$ and the implied shifts in
$S_v$ and $L$ predicted by pure neutron matter are much larger than
the kinetic energy contribution.  If the shifts in $S_v$ and $L$ have
the same sign, as they do for the kinetic energy contribution, the
visible effect in Figure~\ref{correl} would be small.  However, the
shifts could have opposite signs.  More study of fourth-order
corrections in the nuclear symmetry energy would be useful.

\section{Discussion}
\subsection{Neutron Star Radii}

The most direct connection between astrophysical observations and the
nuclear symmetry energy concerns neutron star radii ($R$), which are
highly correlated with neutron star pressures near $n_s$.  \cite{LP01}
found that the radii of neutron stars for masses near 1.4 M$_\odot$
obey a power law relation:
\begin{equation}\label{rp}
R(M)=C(n,M)(p(n)/{\rm MeV~fm}^{-3})^{1/4},
\end{equation}
where $R(M)$ is the radius of a star of mass $M$, $p(n)$ is the
pressure of neutron star matter at density $n$, and $C(n,M)$ is a
constant for a given density and mass. \cite{LP01} found that
$C(n_s,1.4{\rm~M}_\odot)=9.30\pm0.60{\rm~km}$ and $C(2n_s,1.4{\rm~M}_\odot)=5.72\pm0.25{\rm~km}$.  These values were
estimated by averaging the results of 31 disparate equations of state, several of which
are now ruled out because of the maximum mass constraint imposed by
PSR J1614-2230 \citep{demorest10}.  Excluding those ruled-out
equations of state, we find a revised value
%%%%
\be\label{cns}
C(n_s,1.4{\rm~M}_\odot)=9.52\pm0.49{\rm~km},
\ee
%%%%%%%%%%%
with a standard deviation 20\% smaller than previously obtained.
For comparison, at $2n_s$, we find the fractional errors in the revised constant are now smaller by nearly a
factor of 2:
$C(2n_s,1.4{\rm~M}_\odot)=5.68\pm0.14{\rm~km}$.    Although this relation is a tighter fit when applied for densities somewhat in excess of the nuclear
saturation density (e.g., $1.5-2 n_s$), neutron star matter pressure 
cannot be accurately predicted using the symmetry parameters $S_v$ and $L$ alone at these densities.
Hence, we utilize Eq. (\ref{rp}), evaluated at $n_s$, for predictions.

The pressure of pure neutron matter at $n_s$ in the quadratic
  approximation (Eq.~\ref{symm}) is $p(n_s,x=0)=Ln_s/3$.  However, neutron star matter is in
  $\beta$ equilibrium with a finite proton fraction and has a smaller pressure by a factor
\begin{equation}\label{pfac}
f\equiv{p(n_s,x_{\beta s})\over p(n_s,0)}=1-x_{\beta s}\left[4(1-x_{\beta s})-{3S_v\over L}(1-2x_{\beta s})\right].
\end{equation}
$x_{\beta s}$, the proton fraction in $\beta$ equilibrium at the saturation density,
is determined from $S_v$ by the cubic equation
\begin{equation}\label{cubic}
4S_v(1-2x_{\beta s})=\hbar c(3\pi^2n_sx_{\beta s})^{1/3}.
\end{equation}
For the range of values within the overlap region indicated in Fig.~\ref{correl}, $29.0{\rm~MeV}<S_v<32.7{\rm~MeV}$ and $40.5{\rm~MeV}<L<61.9{\rm~MeV}$,
these relations imply $0.035<x_{\beta s}<0.046$ and $0.90<f<0.94$.

To estimate the radius and its error for 1.4 M$_\odot$ stars, we replaced the polygonal overlap region in Fig.~\ref{correl} with an error ellipse such that Monte Carlo
sampling would result in 68\% of the points lying within the polygon.  The ellipse has a semimajor (semiminor) axis of 6.7 (0.55) MeV and an orientation
angle of $8.6^\circ$ from the vertical in the $S_v-L$ plane.  With these specifications, we find that, with Eq. (\ref{cns}), $R_{1.4}=11.9\pm0.7$~km to $1\sigma$ and $R_{1.4}=11.9\pm1.2$~km to $90\%$ confidence.

A study by \cite{Psonis07}, using the parameterized equation of
  state of \cite{PAL} and assuming $S_v=30$ MeV, found a nearly linear
  correlation between $R_{1.4}$ and $L$, $R_{1.4}\simeq8.702+0.070L$
  km, which agrees with Eqs.~(\ref{rp}) and (\ref{cns}) to within
  their errors only for the interval $20{\rm~MeV}<L<60{\rm~MeV}$.  The
  study of \cite{Gearheart11} is only marginally consistent with
  predictions of Eqs.~(\ref{rp}) and (\ref{cns}), perhaps because this study did not
  impose the neutron star maximum mass constraint.

Although more than 30 neutron star masses are now measured to a
reasonable precision~\citep{Lattimer12}, there are no precise
measurements of neutron star radii.  Among the most promising sources
for determining radii are those stars having observed thermal
radiation from their surfaces as well as certain X-ray bursts.

The interpretation of thermal emission has several difficulties,
including uncertainties in distance, atmospheric composition, surface
magnetic field, and interstellar absorption.  Now-quiescent neutron
stars in formerly accreting binaries in globular
clusters~\citep{Heinke06,webb07,guillot11} are promising because it is
believed that recent accretion has both suppressed the surface
magnetic fields and rendered their atmospheres with a nearly pure
hydrogen composition.  In addition, there have been significant
advances in estimating distances to globular clusters.  If these
sources were true blackbodies, measured fluxes and temperatures would
directly yield their apparent angular radii, $R_\infty/D$, where
$R_\infty=R(1-2\beta)^{-1/2}$, $\beta=GM/Rc^2$ and $M$ is the stellar mass. 
These sources imply a wide range of values for $R_\infty$.  A recent study found radii ranging
from 8 km to 23 km \citep{guillot13}, using a low magnetic field, pure
hydrogen atmosphere.  Under the assumption that 5 quiescent sources
would have a common radius, but possibly varying masses,
motivated by the nearly vertical $M-R$ trajectories found by
\cite{SLB10,SLB13} (see below), obtained a joint fit with $R=9.1^{+1.3}_{-1.5}$ km to
90\% confidence, which is inconsistent with our results.  However, this analysis has the theoretically
unlikely result that, even with a common radius, the masses range from 0.86 M$_\odot$ to 2.42
M$_\odot$ \citep{guillot13}.  \cite{Lattimer13} suggest that uncertainties in the amount of interstellar absorption and atmosphere compositions contribute to
the wide predicted ranges of mass and radius.

There are a few non-globular cluster sources having thermal emission, but only
one, RXJ 1856-3954, has a precisely measured
distance~\citep{Walter10}.  \cite{Ho07} modeled it with a carbon-rich
surface with a thin layer of hydrogen.  Including only distance
uncertainties, they found $M\simeq1.33\pm0.09$ M$_\odot$ and
$R\simeq11.9\pm0.8$ km.  However, no confirmation of this assumed
atmosphere composition is available.

Another class of sources, Type I photospheric radius-expansion (PRE)
X-ray bursts~\citep{SB04}, are thought to be
powerful enough to temporarily lift material from neutron star surfaces.  They have observed
maximum fluxes thought to be near their Eddington
limit values, $F_{Edd}=cGM/(\kappa D^2)$ (neglecting redshift), for which
radiation pressure balances gravity ($\kappa$ is the opacity). Combined with angular sizes
determined from the tail of the burst, this permits simultaneous
mass and radius determinations. For these sources, major uncertainties
include distances as well as atmospheric properties. It is also uncertain if
the location of the photosphere, when the maximum flux is observed, is close enough
to the stellar radius to make redshift corrections to $F_{Edd}$ significant. Estimated errors
in $M$ and $R$ are of order 20\% or more for each source.

\cite{SLB10} utilized $M-R$ data from observations of an ensemble of four quiescent and four PRE burst sources
to estimate the parameters for an equation of state whose symmetry energy near the nuclear saturation
density was parameterized by Eq.~(\ref{S20} iii).  Their
Bayesian analysis found, to $1\sigma$,
$28.5{\rm~MeV}\simle S_v\simle34{\rm~MeV}$ and $0.2\simle\gamma\simle0.4$,
which implies $41{\rm~MeV}\simle L\simle51{\rm~MeV}$.  They also found, to
$1\sigma$ for their baseline case, $11.5{\rm~km}\simle
R_{1.4}\simle12.2{\rm~km}$.  Alternatively, utilizing their derived
constraints for $S_v$ and $L$, Eq. (\ref{rp})
predicts $11.0{\rm~km}\simle R_{1.4}\simle12.3{\rm~km}$,
which is quite consistent.  In contrast, studies of PRE bursts by \"Ozel and collaborators
\cite{Ozel10, Ozel12}, deduced radii with $2\sigma$ uncertainties in the range
9--11 km.    Also, \cite{Suleimanov11} proposed that the short PRE bursts considered
by Ozel et al. and Steiner et al.  have systematic uncertainties, such
as disc shadowing, that make them unsuitable for mass and radius
determinations.  They instead proposed the consideration of
longer PRE bursts from which they find significantly larger radii, with estimates
extending beyond 14 km), which necessarily imply larger $L$ values
(100 MeV or larger, according to Eq. (\ref{rp})), which are inconsistent with nuclear
systematics. 
Collectively, these differences suggest that current models are subject to significant systematic
uncertainties involving color correction factors, redshift corrections, and geometry.

A study by \cite{SG12} combined neutron matter calculations and
  astrophysical observations to constrain $L$.  They assumed the
  equation of state in the vicinity of the nuclear saturation density
  was parameterized by an energy density functional with parameters
  taken from the neutron matter results of \cite{GCR12}.  They found,
  for neutron star models both with and without deconfined quark
  matter at high densities, that, to $1\sigma$, $43{\rm~MeV}\simle
  L\simle52{\rm~MeV}$ and $32{\rm MeV}\simle S_v\simle34{\rm~MeV}$, and, to
  $2\sigma$, $39{\rm~MeV}\simle L\simle54{\rm~MeV}$ and $31{\rm~MeV}\simle
  S_v\simle34.5{\rm~MeV}$.  To $1\sigma$, they also inferred a neutron
  star radius range of 10.75--12.25 km for a 1.4 M$_\odot$ neutron
  star, only about 0.35 km smaller, on average, than obtained by \cite{SLB10}.

A more recent study~\citep{SLB13} considered additional sources,
  and in addition considered systematic variations to the fiducial
  analysis in which one or more sources were excluded from the
  Bayesian analysis.  In extreme cases, all PRE X-ray burst sources or
  quiescent globular cluster sources were excluded.  With these
  variations in input data, it was inferred that $43.3{\rm~MeV}\simle
  L\simle66.5$ MeV (Figure~\ref{correl}), to $1\sigma$, somewhat larger on average
  than that deduced by \cite{SLB10} and \cite{SG12}, but very close to
  the range found from experimental results.  In agreement with the
preceding studies, it was found that $S_v$ is relatively poorly constrained.  The $1\sigma$ 
radius range was found to be $10.6{\rm~km}\simle R\simle12.6{\rm~km}$.   The average value
is the same as determined by \cite{SLB10} but the range is broader due to the consideration
of a greater variety of scenarios in an attempt to capture systematic uncertainties in models.
It should be emphasized that neither \cite{SLB10} nor
  \cite{SLB13} assumed an apriori relation between the symmetry
  parameters, despite the fact that nuclear binding energies imply a strong
  correlation exists.

Finally, we note recent analyses~\citep{LMC11} of pulsed light from
rotating, accreting millisecond pulsars {are consistent with a mass-radius relation
similar to that deduced by \cite{SLB10}.

\subsection{Moments of Inertia and the Quadrupole Response}

The moment of inertia depends sensitively on the neutron star radius, scaling approximately as $R^2$.  Hence, constraints to
the neutron star radius will also constrain a neutron star's moment of inertia.  The pulsar binary
PSR 0737 \citep{Lyne04} has detectable spin-orbit coupling and the potential exists for a measurement of 
the most massive component's moment of inertia.  Since the masses in this binary are known to
a high degree of precision,  this presents the possibility of measuring the radius of a neutron star.
\cite{LS05} found that the moment of inertia, for neutron star equations of state capable
of supporting at least 1.6 M$_\odot$, can be well approximated by
\begin{equation}\label{mom}
I=(0.237\pm0.008)MR^2(1+2.84\beta+18.9\beta^4),
\end{equation}
where $\beta=GM/Rc^2$ is the neutron star compactness parameter.
For a 1.4 M$_\odot$ neutron star, this relation, coupled with the radius constraints implied by the experimental
limits to $S_v$ and $L$, predicts $I_{1.4}=71.0\pm7.1 {\rm M}_\odot{\rm~km}^3$.  Another approximation for the moment of inertia~\citep{Link99}, after correction of a typographical error, is
\begin{equation}\label{mom1}
I\simeq{2\over7}MR^2\left(1-{33\over28}\beta\right)^{-1},
\end{equation}
which gives $I_{1.4}=71.4\pm7.2{\rm~M}_\odot{\rm~km}^3$, a nearly identical result.

\cite{FH08} have recently pointed out that tidal effects are
potentially measurable during the late stages of inspiral of a binary
merger involving neutron stars.  Gravitational waves emitted early
during the detection phase (when the orbital frequency becomes larger
than about 100 Hz) are nearly sinusoidal with gradually increasing
frequency and amplitudes.  Tidal forces induce quadrupole moments on
the neutron stars, producing a very small correction which appears as
an accumulated phase shift in the gravitational wave signal.  This
shift will be proportional to a weighted average of the induced
quadrupole polarizabilites ($\lambda_{tid}$, also known as the Love
number) for the individual stars.  Since both neutron stars have the
same equation of state, this weighted average is relatively
insensitive to the binary mass ratio \citep{Hinderer10}.  Since
$\lambda$ scales as $R^5$, it  would provide a sensitive
  measurement of the neutron star radius.  However, the results of
\cite{Hinderer10} and \cite{Postnikov10} show that for radii
restricted to our favored range, $\lambda_{tid}$ is small, making
detection of the tidal signature by Advanced LIGO 
difficult \citep{Hinderer10}.  Much of the difficulty has to do with
degeneracies between the spins of the neutron stars, their rotational
quadrupole moments ($Q_{rot}$) and their tidal Love numbers $\lambda_{tid}$ 
in their
extraction from gravitational wave signals. 

However, \cite{yy2013}
pointed out that nearly EOS-independent relations exist among
$Q_{rot}$, $I$ and $\lambda_{tid}$ which can be used to break these
degeneracies.  In particular, the phenomenological relations
\cite{yy2013} found relating $I$ to $\lambda_{tid}$ and $Q_{rot}$ have interesting
consequences.  These are
\begin{eqnarray}\label{yy}
\bar I\equiv {I/ M^3}&\simeq&\exp\left(1.47+0.817z+1.49z^2+0.287z^3-0.364z^4\right),\cr
\bar Q_{rot}\equiv Q_{rot}/M^5&\simeq&\exp\left(0.194+.936z+.474z^2-4.21z^3+1.23z^4\right),
\end{eqnarray}
where $z=(\ln\bar\lambda_{tid})/10$ and
$\bar\lambda_{tid}=\lambda_{tid}/M^5$.  It is claimed these relations are
accurate to better than a percent for all realistic EOSs and for
neutron star masses greater than about 0.5 M$_\odot$, although they
studied only four EOSs.  We have verified the general validity of this
relation, although we find it holds to slightly lesser accuracy than
claimed.  We find this relation holds to 2\% accuracy for all the
hadronic EOSs described in \cite{LP01} for stars greater than 1
M$_\odot$, except for PS, WFF3 and WFF4.  For the latter EOSs, none of
which satisfy the minimum maximum mass constraint, the error increases
beyond 2\%, but only for$M<1.4 M_\odot$.

These equations imply that, for a 1.4 M$_\odot$ star,
$\bar\lambda_{tid}\simeq72\pm32$ and $\bar
Q_{rot}\simeq3.13\pm0.45$.

These relatively EOS-independent relations can be combined
with Eq.~(\ref{mom}), which itself is accurate to about 3\% for the
same mass range and the same set of EOSs, excluding those that cannot
satisfy the minimum maximum mass constraint imposed by PSR 0737.  This
results in relations relating $\bar\lambda_{tid}$ and $\bar Q_{rot}$ to the
compactness $\beta$ alone.  We determined inverse relationships for Eq. (\ref{yy}) for this purpose:
\begin{eqnarray}\label{yy1}
\bar\lambda_{tid}&\simeq&\exp\left(-4.24+5.96u-3.01u^2+0.722u^3-0.0655u^4\right),\cr
\bar Q_{rot}&\simeq&\exp\left(-1.16-0.700u+1.83u^2-0.593u^3+0.062u^4\right),
\end{eqnarray}
where 
\begin{equation}\label{udef}
u=\ln(\bar I)\simeq\ln\left[0.237\beta^{-2}\left(1+2.84\beta+18.9\beta^4\right)\right].
\end{equation}
We confirmed that these relations are
valid to within about 3\% for masses larger than 1 M$_\odot$ and
$\beta\simle0.3$ for EOSs that satisfiy the minimum maximum mass
constraint.  \cite{Maselli13} points out that a measurement of
$\lambda_{tid}$ to even 60\% accuracy can be used to determine $\beta$
to 10\% percent accuracy, because $\lambda_{tid}\propto R^5$.  This
would translate into a similar accuracy for the radius assuming the
masses can be determined to at least this accuracy.  However, we note
that a simpler expression for $\bar\lambda_{tid}(\beta)$ obtained by
\cite{Maselli13} is considerably less accurate than Eq.~(\ref{yy1}) when
applied to the same suite of EOSs and for the same conditions on mass
and compactness.

 \subsection{Binding Energies}\label{sec:be}

One of the more accurate measurements that can be determined from the neutrino signal from a galactic gravitational collapse supernova is
the neutron star binding energy, assuming the distance to the supernova can also be found.  \cite{LP01} determined a fit to the binding energy (BE) of 
a neutron star, for EOSs which support at least 1.65 M$_\odot$:
\begin{equation}\label{be}
{{\rm BE}\over M} \simeq  (0.60\pm0.05)\beta\left(1-{\beta\over2}\right)^{-1}.
\end{equation}
Using the radius constraints implied by the experimental limits to $S_v$ and $L$ implies BE $\simeq 279^{+35}_{-25}$ B for a 1.4 M$_\odot$ neutron star, where 1 B is $10^{51}$ ergs.
\subsection{Disk Masses and Energetics of Compact Star Mergers}

Black hole-neutron star mergers are among the most likely progenitors
of short gamma-ray bursts \citep{Lee07}, sources of gravitational
radiation (see, for example, \citep{Abadie10, Somiya12}, and sites for
the astrophysical r-process \citep{Lattimer76, Lattimer77}.  Gamma-ray
bursts and r-process production are most likely in the case the
neutron star is tidally disrupted, and radioactive decay in unbound
neutron-rich ejecta could produce an isotropic optical
\citep{Roberts11, Metzger12} or infrared \citep{Kasen13}
transient.  The potential for mass ejection depends on whether or not
tidal effects are strong enough for the neutron star to be disrupted
before reaching the innermost stable circular orbit ($R_{ISCO}$) of
the black hole.  In this case, some disrupted matter remains outside
the black hole for times up to seconds in an accretion disk, a tidal
tail or in unbound matter.  Primarily, the outcome depends on the mass
ratio of the binary ($q=M_{BH}/M_{NS}$), the spin of the black hole
($\chi=a/M_{BH}$), and the neutron star compactness ($\beta$).  Low
mass ratios, high spins, and large neutron star radii are favorable
for large disk masses.  Recently, \cite{Foucart12} has performed a
study of numerical simulations in which these parameters were varied
and formulated a two-parameter model that predicts the disk mass
remaining outside the black hole a few milliseconds following a
merger.  The observed energy of a short gamma-ray burst \citep{Giacomazzo12} and, possibly,
the mass of unbound matter, may be proportional to the disk mass, so
observations of gamma-ray bursts can be compared to models of
population synthesis of compact binaries in support of this
hypothesis.

The fit of \cite{Foucart12} predicts that the disk mass fraction $M_t/M_{NS}$ resulting from a merger varies with $q, \beta$ and $\chi$ according to
\begin{equation}\label{foucart}
{M_t\over M_{NS}}={M^b_{NS}\over M_{NS}}\left[a_1(3q)^{1/3}(1-2\beta)-a_2\beta q{R_{ISCO}c^2\over GM_{BH}}\right],
\end{equation}
where $M^b_{NS}$ is the baryon mass of the neutron star and
$a_1=0.288\pm0.011$ and $a_2=0.148\pm0.007$ are parameters determined
by a fit to merger simulations.  The radius $R_{ISCO}$ of the
innermost stable circular orbit depends on $\chi$ according to
\begin{equation}\label{ISCO2}
\chi={1\over3}\sqrt{R_{ISCO} c^2\over GM_{BH}}\left[4-\sqrt{3{R_{ISCO} c^2\over GM_{BH}}-2}\right],
\end{equation}
valid for either prograde or retrograde orbits.  We also note
that Eq.~(\ref{be}) implies that
\begin{equation}\label{be1}
{M^b_{NS}\over M_{NS}}\simeq1+{3\over5}{\beta\over1-\beta/2}.
\end{equation}
\cite{Foucart12} argues that the major effects of general relativity
are taken into account by the factor $1-2\beta$ and the dependence of
$R_{ISCO}$ on $M_{BH}$ and $\chi$.

\begin{figure}[h]
%\vspace*{-1.1cm}
%\hspace*{-.5cm}
\includegraphics[angle=270,width=18cm,angle=90]{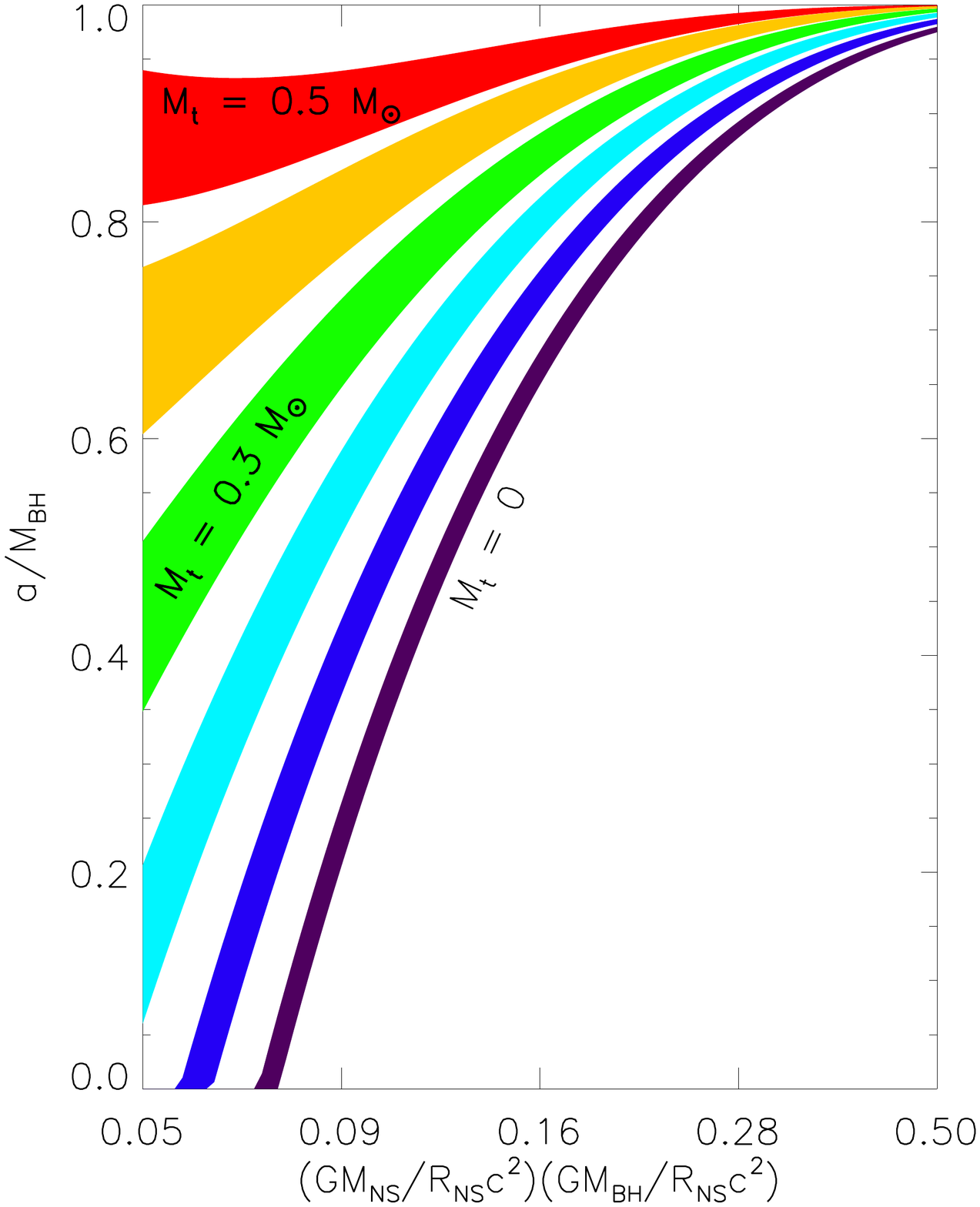}
%\vspace*{-1cm}
\caption{Disk mass in the aftermath of a black hole-neutron star merger as predicted by the model of \cite{Foucart12}, from 0 to 0.5 M$_\odot$, in increments of 0.1 M$_\odot$.  The width of the bands represent the effects of varying $R$ and $M_{NS}$ as discussed in the text.  Note that the $x$-axis is logarithmic.}
\label{bhns}
\end{figure}

Eq.~(\ref{foucart}) can be expressed in the form
$\chi=\chi(q,\beta,M_t, M_{NS})$.  \cite{Foucart12} presented results
as a series of figures illustrating the ratio of disk mass to
$M^b_{NS}$ as a function of the three variables $q, \beta$ and $\chi$.
This four-dimensional function can be reduced to essentially two
dimensions by employing experimental restrictions to the nuclear
symmetry energy which confine neutron star radii to small ranges.
Given that the range of $L$ values deduced from experiments is nearly the
same as that inferred from astrophysical modeling, we use the baseline results
from \cite{SLB10,SLB13} that $11<R<12.5$ km is valid not only
for 1.4 M$_\odot$ neutron stars, but also for stars in the
range 1.2--1.8 $M_\odot$. We further restrict the likely values of $M_{NS}$ to this range.
With these constraints, $0.14\simle\beta\simle0.24$ and  Eq.~(\ref{foucart}) therefore implies a small range of $\chi$
values for particular choices of $q$ and $M_t$.  As a result, the
series of diagrams shown in \cite{Foucart12} can be distilled into a
single picture (Fig. \ref{bhns}).  We find the resulting width of
$\chi$ bands corresponding to given $M_t$ values is minimized if the
abscissa is chosen to be $q\beta^2$ (justified below). Nevertheless, due to the
restricted range of $\beta$, the abscissa
is still dominated by its $q$ dependence.  Generally, the bottom (top)
of each $\chi$ band in Fig. \ref{bhns} corresponds to $M_{NS}=1.2 (1.8) M_\odot$ and
$R=12.5 (11)$ km.  The region to the right of the $M_t=0$ band
corresponds to the situation that tidal disruption occurs within the
ISCO so that no disk is able to form and no short gamma-ray burst will be observed.  It is readily seen that the
disk mass increases rapidly with black hole spin, and that maximally
spinning holes of virtually any stellar mass will form disks after
merging.  The disk mass also increases with decreasing $q$ or $\beta$.
This behavior is easily understood if Eq.~(\ref{foucart}) is rewritten
as
\begin{equation}\label{foucart1}
r\equiv{R_{ISCO}c^2\over GM_{BH}}={a_1(q\beta^2)^{1/3}\beta^{1/3}(1-2\beta)-(M_t/R)(M_{NS}/M_{NS}^b)\over a_2q\beta^2}
%M_t={M^b_{NS}\over M_{NS}}R\left[a_1(3q\beta^2)^{1/3}\beta^{1/3}(1-2\beta)-a_2\beta^2 q\right],
\end{equation}
We note that $0.32\simle\beta^{1/3}(1-2\beta)\simle0.37$ and
$0.86\simle M_{NS}/M_{NS}^b\simle0.92$ are weak functions of $\beta$ in the
relevant range, so that $r$ is largely a function of $M_t$ and
$q\beta^2$ alone.  Also recall that $r$ is a decreasing function of
$\chi=a/M_{BH}$.  Clearly, $r$ decreases and $a/M_{BH}$ increases with
increasing $q^2\beta$ for small to moderate values of $M_t$.  However,
for large enough values of $M_t$, of order 0.5 M$_\odot$ or greater,
the numerator of Eq.~(\ref{foucart1}) vanishes for small values of
$q^2\beta$ so that $\chi\approx1$ for all values of the abscissa.

A further consequence of Eq.~(\ref{foucart}) is that the mass of the
disk and, presumably, the unbound matter will be sensitive to
the neutron star compactness and expected to decrease with increasing $\beta$.
  
The most energetic observed short gamma-ray bursts imply disk masses
of order 0.5 M$_\odot$ \citep{Giacomazzo12}, which can only be
accommodated for black hole-neutron star systems with $\chi\simeq1$.
Alternatively, \cite{Giacomazzo12} suggest that binary neutron star mergers 
could more likely produce large disk masses.  They estimated from general
relativistic simulations using idealized equations of state that
\begin{equation}\label{gia}
M_t\approx3(1-q)(1+q-M_{tot}/M_{max})\simeq3(1-q^2)(1-M_1/M_{max}),
\end{equation}
where $q<1$ is the binary mass ratio, $M_{tot}$ is the total
gravitational mass of the two neutron stars, and $M_{max}$ is the
maximum mass of an isolated neutron star.  This formula predicts that
a disk mass of order 0.5 M$_\odot$ could be produced if $q$ is
significantly different from 1 and if the larger neutron star is considerably
less massive than the maximum mass.  However, 
Eq.~(\ref{gia}) is suspect because it is based on simulations
employing idealized fluid EOSs, which have unphysical $M-R$ relations;
in addition, the dependence on neutron star compactness was not
considered.  It can be expected, in particular, that the disk mass
should decrease with compactness as was already noted for black
hole-neutron star mergers above.  In fact, this inverse behavior is
observed for the mass of unbound matter in the extensive set of
general relativistic simulations employing realistic EOSs 
produced by \cite{Bauswein13}.

\subsection{The Core-Crust Boundary and the Crust
  Thicknesses}\label{sec:crust}
In addition to neutron star radii and moments of inertia, several
properties of the neutron star crust are also impacted by the nuclear
symmetry energy.  The most important effect is the location of the
core-crust boundary, where a phase transition separates heterogeneous
matter containing nuclei and a neutron fluid, and homogeneous, uniform
nuclear matter.  It has been demonstrated that the thickness of the
crust, $\Delta R$, defined as the distance between neutron drip
density and the core-crust transition density, depends primarily on
the stellar radius and the neutron chemical potential at the
core-crust boundary~\citep{LP07}:
\begin{equation}\label{dr}
\Delta R/R\simeq ({\cal H}-1)\left[{\cal H}\left(1-2\beta\right)^{-1}-1\right]^{-1}.
\end{equation}
In this expression, ${\cal H}=\exp[2(\mu_{nt}-\mu_{n0})/m_bc^2]$,
where $\mu_{nt}$ and $\mu_{n0}\simeq-9$ MeV are the neutron chemical
potential at the core-crust transition and at the stellar surface,
respectively.  The crust thickness, together with its transport
properties, determines the thermal relaxation time of the crust, which
is important in in the cooling of young neutron
stars~\citep{yakovlev04}, such as the recently detected rapid cooling
of Cassiopeia A~\citep{Heinke10,Page11} and in the cooling of soft
X-ray transients \citep{PR12}.

A related quantity is the fraction of the star's moment of inertia
residing in the crust, $\Delta I/I$, which depends primarily on the
stellar mass and radius and the pressure at the core-crust boundary,
and scales as $p_tR^4M^{-2}$~\citep{Link99}. The standard model for
pulsar glitches holds that they are due to the neutron superfluid in
the star's crust.  In this case, the observed glitch rates and
magnitudes for the Vela pulsar imply that 
%%%%%%%%
\be\label{dii}
\Delta I/I\simge0.016,
\ee
%%%%%%%%%%%%%%%%
and similar or smaller values are found for other glitching
pulsars~\citep{Andersson12}.

An approximate method of determining the location of the core-crust
boundary is by considering the stability of a homogeneous fluid of
nucleons in beta equilibrium.  \cite{BBP} showed that small density
fluctuations in an otherwise uniform fluid lead to instability when
\begin{equation}\label{df}
\mu_{pp}-\mu_{np}^2\mu_{nn}^{-1}+4e\sqrt{\pi\eta}-4\eta\alpha(9\pi n_p^2)^{1/3}=0,
\end{equation}
where $\mu_{ij}=\partial\mu_i/\partial n_j$ and $\mu_i$ and $n_i$ are
the chemical potential and number density, respectively, of neutrons,
protons or electrons.  $\alpha\simeq1/137$ is the fine-structure
constant. $\beta$ is determined by density gradient terms in the
nuclear Hamiltonian and can be approximated by
$\eta=D[1-4\mu_{np}/\mu_{nn}+(\mu_{np}/\mu_{nn})^2]$ where $D\simeq100$ MeV fm$^{5}$~\citep{HLPS13}.  It
has been assumed that electrons are relativistic and degenerate.
The chemical potentials and derivatives in Eq.~(\ref{df}) are equivalent to
\begin{eqnarray}\label{df1}
\mu_n&=&{\partial(ne)\over\partial n}-x{\partial e\over\partial x},\qquad \mu_p={\partial(ne)\over\partial n}+(1-x){\partial e\over\partial x},\cr
\mu_{nn}\mu_{pp}-\mu_{np}^2&=&{1\over n}{\partial^2(ne)\over\partial n^2}{\partial^2e\over\partial x^2}-\left({\partial^2e\over\partial n\partial x}\right)^2,\cr
\mu_{np}&=&{\partial^2(en)\over\partial n^2}+(1-2x){\partial^2e\over\partial n\partial x}-{x(1-x)\over n}{\partial^2e\over\partial x^2},\cr
\mu_{nn}&=&{\partial^2(en)\over\partial n^2}-2x{\partial^2e\over\partial n\partial x}+{x^2\over n}{\partial^2e\over\partial x^2}.
\end{eqnarray}
Obviously, the functional form of the energy per baryon $e(n,x)$ will
influence the core-crust transition density $u_t=n_t/n_s$ determined
by Eq.~(\ref{df}).  We assume a quadratic symmetry energy with
$S_2(u)$ given by Eq.~(\ref{S20}), and two different functional forms
for the symmetric matter energy,
\begin{equation}\label{funform}
\begin{array}{rll}
  e(u,1/2)&=(3T_0/5)u^{2/3}[1+cu]+a_0u+b_0u^{4/3},\qquad &(i)\cr
  e(u,1/2)&=-B+(K/18)(u-1)^2+k(u-1)^3+d_0(u-1)^4,\qquad &(ii)
\end{array}
\end{equation}
where
\begin{equation}\label{coef1}
a_0=-\left[4B+{6T_0\over5}\left(1-{c\over2}\right)\right],\qquad
b_0={3T_0\over5}(1-2c)+3B,\qquad d_0=B-{K\over18}+k.
\end{equation}
In the above, we assume $B=16$ MeV is the bulk binding energy at $n_s$
and $K=240$ MeV is the incompressibility.  The coefficient $d_0$
ensures the energy per particle vanishes at zero density,
$e(0,1/2)=0$.  With these two forms, and the five associated with
Eq.~(\ref{S20}), we have 10 different expressions for modeling nuclear
matter.  Excluding variations utilizing Eq. (\ref{S20} iii), we found
that the energy and pressure predicted for pure neutron matter can be
adequately approximated, as shown in Fig.~\ref{nm}, for the density
range of interest for determining the core-crust transition densities.
In $\beta-$equilibrium, at these densities, pure neutron matter is
very similar to neutron star matter since proton fractions are of
order 1\%.

\begin{figure}[h]
%\vspace*{-.8cm}
\hspace*{-1.3cm}
\includegraphics[angle=90,width=14cm,angle=90]{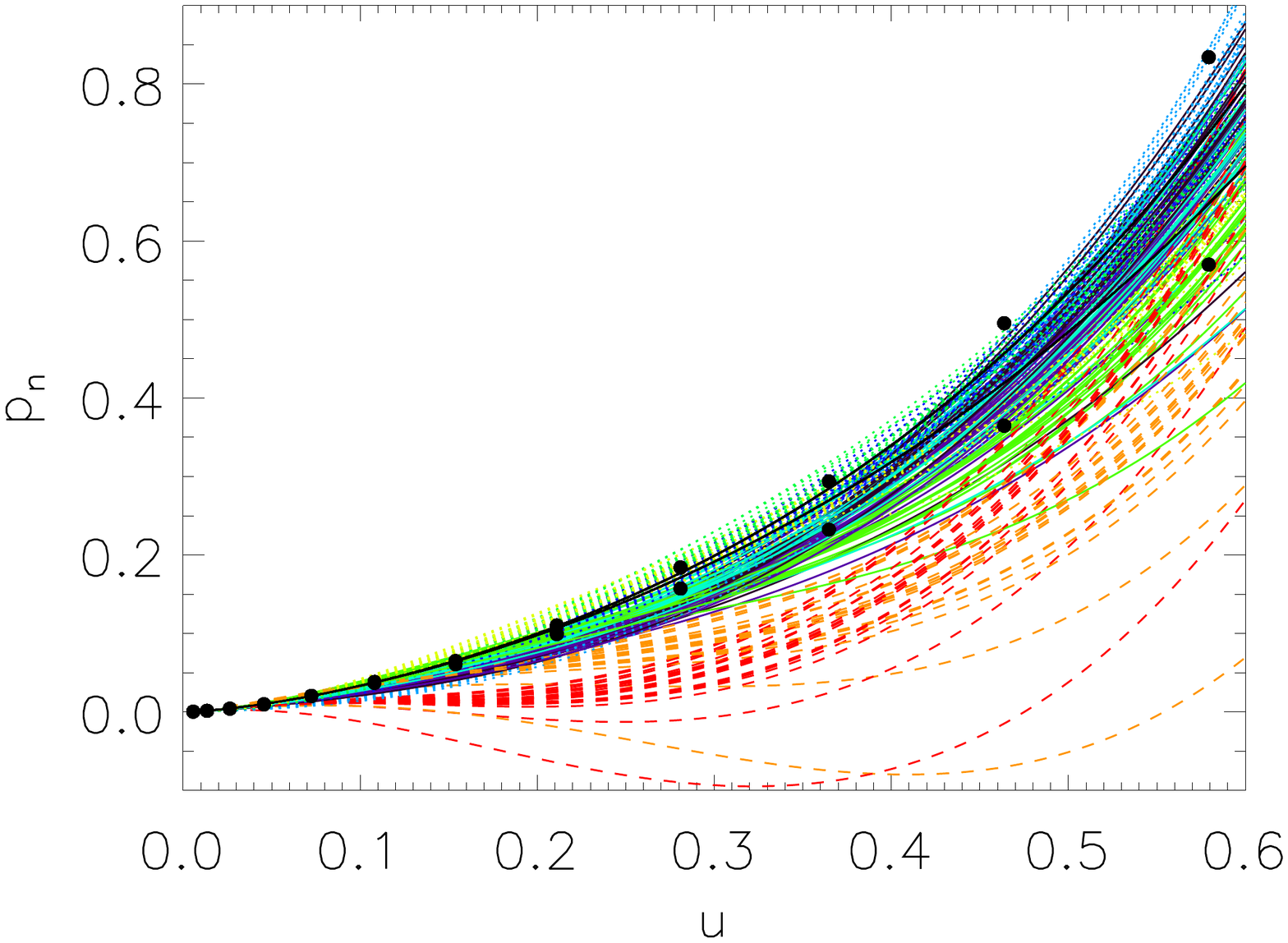}
%\vspace*{-1.5cm}
\caption{Pressures of pure neutron matter as determined from
  Eq.~(\ref{funform}).  Solid, dotted and dashed lines correspond to
  forms i, ii and iii, respectively, for $S_2(u)$, randomly choosing
  force parameters bounded by experimental constraints on $S_v$ and
  $L$ as discussed in the text.  Filled circles indicate bounds for neutron
  matter pressures from \citep{HLPS10}.}
\label{nm}
\end{figure}

Choosing values for $S_v$ and $L$ within the overlap region in
Fig.~\ref{correl}, as previously described, the above energy
expressions can be used to find the core-crust transition density,
pressure and neutron chemical potential, $u_t$, $p_t$, and ${\cal
  H}_t$, respectively.  For each EOS model, $u_t$ ($p_t$ and ${\cal
  H}_t$) decreases (increase) with $L$ or $S_v$ with narrow spreads.
When all possible variations are considered, the values of these
quantities, and in particular $p_t$, are poorly constrained.  However,
limiting $S_2$ to the forms (i) and (ii) of Eq.~(\ref{S20}), supported
by neutron matter studies, results in relatively tight constraints.
Monte Carlo sampling to 90\% confidence limits the transition density
to $0.49\simle u_t\simle0.60$, the pressure to
$0.40{\rm~MeV~fm}^{-3}\simle p_t\simle0.60$ MeV fm$^{-3}$ and ${\cal
  H}$ to $1.043\simle{\cal H}\simle1.053$.  This represents a
considerable improvement over the analysis of \cite{LP01} which
considered interactions leading to $0.2{\rm~MeV~fm}^{-3}\simle
p_t\simle0.65{\rm~MeV~fm}^{-3}$.

In comparison, \cite{HLPS13}, working directly from the properties of
pure neutron matter found $0.48\simle u_t\simle 0.56$ and
$0.41{\rm~MeV~fm}^{-3}\simle p_t\simle0.47{\rm~MeV~fm}^{-3}$.  Both
quantities are within the boundaries of our determinations, consistent
with the relative $S_v$ and $L$ values implied by neutron matter
studies relative to experimental constraints (Fig.~\ref{correl}).
\cite{HLPS13} also found that $u_t$ decreased and $p_t$ increased
linearly with increasing $L$ or $S_v$ values.

The crust thickness of a 1.4 M$_\odot$ star, using Eq.~(\ref{dr}), the
experimental constraints to $S_v$ and $L$, forms (i) and (ii) of
Eq.~(\ref{S20}), and Monte Carlo sampling, can consequently be limited
to $\Delta R/R=0.082\pm0.008$, or $\Delta R=0.96^{+0.17}_{-0.13}$ km,
with $1\sigma$ errors.

The fraction of the moment of inertia of the neutron star crust
depends primarily on the mass and radius of the star as well as $p_t$
and, to a lesser extent, $u_t$ or $h_t$.  
\cite{LP07} showed that this fraction is well approximated by
\begin{equation}\label{di}
{\Delta I\over I}\simeq{8\pi p_tR^4\over3GM^2}\left[{MR^2\over I}-2\beta\right]e^{-4.8\Delta R/R}.
\end{equation}
Applying the experimental constraints to $S_v$ and $L$, and using
Eq. (\ref{mom}) together with Monte Carlo sampling, we find $\Delta
I/I=0.037^{+0.012}_{-0.007}$ for a 1.4 M$_\odot$ star.
%\cite{Link99} showed that this fraction is
%\begin{equation}\label{di}
%{\Delta I\over I}\simeq{28\pi p_tR^4\over3GM^2}
%\left[\left(1-{2GM\over Rc^2}\right)^{-1}+{8p_t\over u_tn_sm_b}\left(5.5-{2GM\over Rc^2}\right)\right]^{-1},
%\end{equation}
%where $I$ is the star's moment of inertia.  Applying the experimental
%constraints on $S_v$ and $L$, which limit $R$, $p_t$ and $u_t$ as
%described above, we find for a 1.4 M$_\odot$ star that $\Delta
%I/I\simeq0.066\pm0.021$.  
For comparison, the approximation suggested by \cite{Link99}, in which the exponential factor
in Eq.~(\ref{di}) is replaced by
\begin{equation}\label{dlink}
\left[1+{2p_t\over n_sm_bu_t}\left({Rc^2\over GM}-2\right)\left({Rc^2\over GM}+7\right)\right]^{-1},
\end{equation}
and $I$ by Eq. (\ref{mom1}), yields 
$[MR^2/I-2\beta]=(7/2)[1-7\beta/4]$ and $\Delta I/I=0.042^{+0.016}_{-0.008}$ for a 1.4 M$_\odot$ star.

The inferred values of $\Delta I/I$ from observations, as discussed above, appears to be consistent with this estimate.  However, this
explanation has been called into doubt~\citep{Andersson12, Chamel12} due to
entrainment of superfluid neutrons in the crust, which raises their
effective masses to several times their bare mass.  As
a consequence, the inferred lower limit to $\Delta I/I$ would have to be larger
by a similar factor, so that 
%%%%%%%%
\be\label{dii1}
\Delta I/I\simge 0.07
\ee
%%%%%%%%%%%%%
would be required to explain glitches.  Only very low-mass neutron
stars, with $M\simle1.2$ M$_\odot$, could satisfy this criterion, calling the standard model for glitches into question.

\cite{Link13} has pointed out, however, that $I$ in Eqs. (\ref{dii})
and (\ref{dii1}) does not necessarily represent the total moment of
inertia of the star.  Although it is usually assumed that the crust
couples to the entire core of the star, this need not be the case.
\cite{Link13} argues that the timescale for the outer core to decouple
from the crust is of the order of weeks to years, and could correspond
to the observed post-glitch relaxation timescale, which may simply
represent the dynamical recovery of the outer core.  If, for example,
the crust couples to a fraction of the core, the inferred value of
$\Delta I/I$ is decreased by this fraction, perhaps a factor of 2 or
more. If this is the case, observations would regain consistency with
theoretical expectations for 1.4 M$_\odot$, and, perhaps, more
massive, stars.

\subsection{Composition of the Neutron Star Interior}

The restrictions on the density dependence of the symmetry energy from
experiments and neutron matter theory expectations not only
establishes important constraints on neutron star radii, but, coupled
with the discovery of a nearly 2 M$_\odot$ neutron star
\citep{demorest10}, considerably reduces uncertainties in predictions
for the properties of high-density matter.   The results of
  \cite{SLB10,SLB13} suggest that even at $6n_s$ the pressure can be
  estimated to within a $1\sigma$ error of about 35\%.  The maximum
mass constraint has also been used to limit the maximum strangeness
(e.g., hyperon) fraction in a neutron star interior:
\cite{Weissenborn12} estimates this to be less than 20\% even if the
maximum mass of a purely hadronic star is about 3.2 M$_\odot$, which
is the largest mass allowed by causality if a 1.4 M$_\odot$ star has
$R\le12$ km~\citep{Lattimer12}.  The strangeness content would be
smaller for less stiff hadronic EOSs and smaller maximum masses.  This
restriction in the  strangeness content of neutron stars is
consistent with studies using the chiral quark-meson
coupling~\citep{Katayama12} and relativistic
mean-field~\citep{Bednarek12} models.

Furthermore, many studies rule
out even the possibility of hybrid stars (those with a mixed phase of quarks and
hadrons, e.g.,~\cite{Dexheimer13}).  The study of \cite{Endo11} found no
stable hybrid stars with maximum masses greater than 1.8 M$_\odot$.
  In general, the phase boundary
can be modeled with either a Maxwell or Gibbs construction, but if the
surface tension of quark matter is sufficiently large, Gibbs
transitions are not allowed.
Many studies indicate that if the phase transition is of the Maxwell
type, the onset of
quarks immediately destabilizes the star so that the existence of a
hybrid star is not possible.   

However, some studies do permit hybrid stars that can satisfy the
minimum maximum mass constraint.  \cite{Lenzi12} find stable hybrid
configurations using the Nambu-Jona-Lasinio (NJL) model with the
relatively stiff (i.e., large values of $L$) GM1 and TM1 nucleon
forces.  However, when hyperons are included, the ranges of parameter
space allowing stable hybrid stars is severely restricted, and the
maximum mass can barely exceed 2 M$_\odot$.  However, they did not
investigate stability using nucleon forces with values of $L$ in the
range constrained by experiments, which lowers the permitted maximum
masses.  \cite{Weissenborn11} has shown that the existence of hybrid
quark stars with a mixed phase and having a sufficiently large maximum
mass is strongly dependent upon the assumed hadronic EOS, and is
limited to small, specific ranges of quark matter parameters. 
  They found a marked sensitivity to the stiffness of the equation of
  state as determined by the radii of moderate (1.4 M$_\odot$) stars,
  so that less stiff equations of state lead to smaller permitted
  parameter ranges for stable hybrid stars.  Nevertheless,
  \cite{Weissenborn11} only considered hadronic interactions with
relatively large values of $L$ ($>100$ MeV).  \cite{Alford13}
  also explored the effects of phase transitions in hybrid stars, but
  also included the dependence on $L$ in their study.  They
  demonstrated that $L$ values in the experimentally preferred range
  severely restricts models capable of satisfying the observed minimum
  maximum mass constraint.  Their models are parameterized by the
  transition density $n_{crit}$, the magnitude of the phase
  transition, $\Delta (ne)=\mu\Delta n$, and the sound speed $c_s^2$
  which is assumed constant for $n>n_{crit}$. For example, with
  $\Delta n=0$, $n_t=2n_s$ and $c_s^2=c^2$, $L\simeq40$ MeV could
  still result in maximum masses of order 2.2 M$_\odot$.  However,
  increasing $\Delta(ne)$ to $0.2(ne)_{crit}$ and $n_{crit}$ to 4$n_s$
  reduces the maximum mass to less than 2 M$_\odot$.  And if
  $c_s^2=1/3$, the value expected asymptotically for quark matter, no
  value of $\Delta(ne)$ can satisfy the maximum mass constraint for
  $n_{crit}\simge1.5n_s$.

\section{Conclusions} 
Our results indicate that there is emerging concordance among
experimental, theoretical and observational studies for values of the
nuclear symmetry energy parameters and neutron star radii.  We showed
that existing estimates of the symmetry parameters from nuclear
experiments measuring neutron skin thicknesses, dipole
polarizabilities and giant dipole resonances, and from heavy ion
collisions, can be made more precise by including the powerful
constraints available from nuclear mass fitting.  We obtained joint
limits to the symmetry parameters from combining the experimental
results: 29.0 MeV $<S_v<$ 32.7 MeV and 40.5 MeV $<L<$ 61.9 MeV.
  From theoretical studies of the correlation between neutron star
  pressures near the nuclear saturation density and neutron star
  radii, we determine that $R_{1.4}\simeq11.9\pm1.2$ km to 90\% confidence.
  Independently, information from theoretical neutron matter
studies and observations of neutron star masses and radii
restrict the neutron star radius for stars in the vicinity of 1.4
M$_\odot$ to a somewhat smaller range, 11 km $<R_{1.4}<$ 12.5 km
(see \citep{SLB10,HLPS10,GCR12}).  This reinforces the validity of our
  experimental constraints for the symmetry energy and suggests that
  uncertainties in the properties of matter near the nuclear
  saturation density, and in the estimated neutron star radius, can be
  further reduced by future experiments and observations.

  There have been other recent studies with similar goals.  Those
  pertaining to observations of neutron star masses and radii were
  summarized in the preceding section.  In addition, numerous studies
  have utilized experimental studies to constrain $S_v$ and $L$.  Most
  of these studies focused on the implications of single classes of
  experiments.  

  \cite{Newton11} focused on studies of neutron skin thicknesses,
  isospin diffusion, isobaric analog states, pygmy dipole resonances,
  optical model potentials, and droplet model predictions.  They also
  included some aspects of the correlations resulting from nuclear
  binding energies that we have discussed.  However, those
  correlations did not include error bars in several cases and were
  based on specific force models rather than the generic correlation
  we attempted to establish.  They also discussed implications for the
  neutron star radius, moment of inertia, the core-crust transition
  density and pressure, the inner crust-mantle transition density and
  the fractional crust thickness and moment of inertia.  They found
  that the crust-core transition density decreases with the assumed
  value of $L$ and that the transition pressure increases with $L$.
  The possible ranges of $u_t$ and $p_t$ they found, however, were
  $0.50\simle u_t\simle0.75$ and
  $0.35{\rm~MeV~fm}^{-3}\simle0.85{\rm~MeV-fm}^{-3}$, respectively,
  both much wider than the ranges we inferred.  They discussed
  implications for the possible extent of pasta phases, a subject we
  did not explore.  Their results for the fractional crust thickness
  and moment of inertia also showed broader ranges than ours.
  Finally, they considered the fundamental and overtone frequencies
  from the spectrum of torsional shear oscillations in the stellar
  crust.  However, the comparison with data is difficult because of
  superfluid entrainment effects and assumptions concerning the phase
  of the mantle, as well as the extent of coupling of crustal
  oscillations with the core.  Due to these extensive uncertainties,
  we have chosen not to consider these frequencies.  Finally,
  \cite{Newton11} considered the $L$ dependence of the maximum
  quadrupole ellipticity of the crust which sets an upper limit to the
  signal that gravitational wave detectors can observe.

\cite{Tsang12} also recently undertook a comprehensive analysis of
heavy-ion collisions, isobaric analog states, nuclear binding
energies, pygmy dipole resonances, neutron skin thicknesses, electric
dipole polarizabilities, neutron matter, and microscopic nuclear
models in an attempt to constrain the symmetry energy parameters.
However, their analysis differs significantly from our study.  First,
their discussion of nuclear binding energies was limited to the
results of the finite range droplet model FRDM12\citep{Moller12} and
did not include the correlation of symmetry parameters established
here and studied in, for example, \cite{Kortelainen10} or
\cite{Oyamatsu03}.  The best fit values for $S_v$ and $L$ for the
UNEDF0, UNEDF1, FRDM90, FRTF or LD models we show in Figure
\ref{massfit} do not lie within the error region they indicated for
FRDM12, showing that the size of this error region is substantially
underestimated.  Second, \cite{Tsang12} determined a weighted average
for $r_{np}^{208}$, namely $0.19\pm0.03$ fm, which is slightly larger
than we assume.  However, they used this result to infer a range of
values for $L$ rather than incorporating the $S_v-L$ correlation that
\cite{Chen10} established (and we confirmed) for the neutron skin
thickness to more tightly constrain both symmetry parameters.  Third,
\cite{Tsang12} established correlations between $S_v$ and $L$ from the
neutron matter studies of \cite{GCR12} and \cite{HLPS10} that are
virtually identical to ours, but they did not derive corresponding
error regions.  Fourth, \cite{Tsang12} utilized a detailed study
\citep{Dutra12} of 240 Skyrme interactions, limited to those 5 that
survived 11 macroscopic and 4 microscopic constraints, to infer a
correlation between $S_v$ and $L$. In a sense, this correlation is
analogous to the binding energy constraint we emphasized here, but
does not have the same properties.  Finally, Tsang et al. did not use
measurements of the giant dipole resonance to provide additional
constraints to $S_v$ and $L$.  Overall, they concluded that
experiments constrain the symmetry parameters to a range of values
centered on $S_v\approx32.5$ MeV and $L\approx70$ MeV, but did not
quantify those ranges.  Both central values are larger than our
preferred values.  This is partly due to our rejection of isobaric
analog state and pygmy dipole resonance estimates, and partly to our
utilization of a more comprehensive study of the constraints available
from fitting nuclear masses.

In addition to the role played by the nuclear symmetry energy in
crustal properties, the pulsar glitch phenomenon, and in neutron star
cooling, its influence on neutron star radii has other far-reaching
consequences.  The neutron star radius will influence the neutrino
emission during supernovae and proto-neutron star formation, and
therefore has ramifications for the supernova mechanism and r-process
production in the aftermath of supernovae.  In addition, the tidal
quadrupole polarizability of neutron stars depends on $R^5$ which has
important ramifications for gravitational wave signals.  The radius
also strongly affects the quantity of mass left in disks after a
merger event involving neutron stars, which will leave signatures in
the energetics of short gamma-ray bursts and also electromagnetic
signals (afterglows) due to r-process nucleosynthsis in ejected,
unbound matter. 

\begin{acknowledgments}
This work was initiated by the U.S. DOE-funded Topical Collaboration at Los Alamos, and was partially supported by the U.S.~DOE grant
DE-AC02-87ER40317.
\end{acknowledgments}

\end{document}